\documentclass[sigconf, nonacm, 9pt]{acmart}

\usepackage[english]{babel}
\usepackage{blindtext}
\usepackage{xurl}

\usepackage{caption}
\usepackage{comment}
\usepackage{tikz}
\usepackage{amsmath}

\usepackage{subcaption}
\usepackage{graphicx}

\usepackage{enumitem}
\usepackage{pbox}
\usepackage{multirow}
\usepackage{tabularx}
\usepackage{makecell}
\usepackage{mathtools}
\usepackage{caption}
\usepackage{float}
\usepackage{booktabs}
\usepackage{xurl}
\usepackage{gensymb}

\usepackage{hyperref}
\usepackage{graphicx}

\newcommand{\parab}[1]{\vspace{0.05in}\noindent\textbf{#1}}

\newcolumntype{L}[1]{>{\raggedright\let\newline\\\arraybackslash\hspace{0pt}}m{#1}}
\newcolumntype{C}[1]{>{\centering\let\newline\\\arraybackslash\hspace{0pt}}m{#1}}
\newcolumntype{R}[1]{>{\raggedleft\let\newline\\\arraybackslash\hspace{0pt}}m{#1}}

\setcopyright{none}

\renewcommand\footnotetextcopyrightpermission[1]{}
\begin{document}

\title[]{A Deep Dive into the Impact of Solar Storms \\ on LEO Satellite Networks}

\author{Eunju Kang}
 \orcid{0009-0005-3591-5758}
 \affiliation{%
   \institution{University of California, Irvine}
   \country{USA}
 }
 \email{eunjuk2@uci.edu}

 \author{Alagappan Ramanathan}
 \orcid{0009-0003-3293-1790}
 \affiliation{%
   \institution{University of California, Irvine}
   \country{USA}
 }
 \email{alagappr@uci.edu}

 \author{Sangeetha Abdu Jyothi}
 \orcid{0009-0000-0503-4478}
 \affiliation{%
   \institution{University of California, Irvine}
   \country{USA}
   }
   \email{sangeetha.aj@uci.edu}

 \renewcommand{\shortauthors}{Eunju Kang, Alagappan Ramanathan and Sangeetha Abdu Jyothi}

\begin{abstract}
Low Earth Orbit (LEO) satellite networks are an important part of the global communication infrastructure today. Despite ongoing efforts to improve their resilience, they remain vulnerable to component damage and deorbiting under harsh space weather conditions. Prior work identified a modest but noticeable impact on LEO satellite network performance during solar storms, typically manifesting as an immediate rise in packet loss and a sustained increase in round-trip time (RTT). However, these studies offer only coarse-grained insights and do not capture the nuanced spatial and temporal patterns of disruption across the LEO network. 

In this paper, we conduct a deep dive into the impact of solar storms on LEO satellite communications. By localizing the impact of increased atmospheric drag at the level of individual satellites and orbits, we reveal significant heterogeneity in how different parts of the network are affected. We find that the degree of performance degradation varies significantly across geographic regions, depending on satellite positioning during the storm. Specifically, we find that \textit{(i)} not all satellite orbits are equally vulnerable, \textit{(ii)} within a given orbit, certain satellites experience disproportionate impact depending on their position relative to geomagnetic conditions, and \textit{(iii)} autonomous maneuvering of satellites might be a cause of the sustained increase in RTT. Our findings uncover previously overlooked patterns of vulnerability in LEO satellite constellations and highlight the need for more adaptive, region-aware mitigation strategies to address space weather-induced network disruptions.

\end{abstract}
\maketitle
\section{Introduction}

Low Earth Orbit (LEO) satellite constellations are rapidly transforming global connectivity. With their ability to provide low-latency, wide-area coverage, LEO networks have become foundational to applications ranging from consumer Internet access to disaster response and maritime communications. Major initiatives such as Starlink~\cite{Starlink} and OneWeb~\cite{OneWeb} have already deployed thousands of satellites, with plans for tens of thousands more. As dependence on LEO-based communication grows, ensuring its robustness becomes increasingly critical.

A major threat to LEO satellite operations is space weather. Satellites are directly exposed to extreme space weather events, such as solar flares and Coronal Mass Ejections (CMEs) \cite{sigcomm-solarstorm}. These events, driven by increased solar activity, can elevate atmospheric drag at orbital altitudes, stressing hardware systems and affecting satellite trajectories~\cite{cosmicDance,past-work-1,past-work-2,past-work-3,past-work-4,past-work-5, sat-resilience-db-1, sat-resilience-db-2}. While significant engineering efforts have gone into improving satellite resilience, operational data from recent storms reveal persistent vulnerabilities, including network performance degradation and, in extreme cases, satellite deorbiting~\cite{spacex-impact-1}.

Prior work has shown that solar storms can cause modest yet measurable degradation in communication performance, typically manifesting as increased packet loss and elevated Round-Trip Time (RTT)~\cite{leonet24-solarstorm}. However, past work tends to view the LEO network in aggregate, offering only coarse-grained insights. They do not account for the complex spatial and temporal dynamics of how different satellites and regions are affected during a storm. This gap limits our ability to design targeted mitigation strategies and adaptive communication protocols.

In this paper, we present a fine-grained analysis of solar storm impacts on LEO satellite networks. Using real-world network measurement data from RIPE probes connected to the Starlink network (AS 14593) and publicly available satellite position information across four solar storms in 2024, we investigate how atmospheric drag from geomagnetic disturbances affects different parts of the LEO constellation, and in turn, network performance. 

Our study reveals significant heterogeneity in the impact based on the time and location of the storm. \textit{(i) Geographic variation}: There are variations in network performance degradation across regions in terms of the time of peak impact. \textit{(ii) Orbital disparities}: Not all orbital planes experience the same level of disruption during a storm. The extent of impact depends on the orientation of the orbit with respect to the sun during storm impact. \textit{(iii) Location-based variations}: We recognize three broad location-based categories for highly impacted satellites. We identify relatively large altitude changes for satellites at high latitudes, those over the South Atlantic Anomaly (SAA) region~\cite{saa-2}, and those facing the sun during the time of solar storm impact. Additionally, we also uncover patterns in highly affected satellites across consecutive storm days, showing the propagation of impact across neighboring satellites.

By uncovering spatial patterns of vulnerability, our work provides actionable insights to improve the robustness of LEO communication systems. We also discuss implications for adaptive routing and long-term constellation design in the face of increasing solar activity.

\section{Background}
We provide a brief overview of prior work on the analysis of satellite network performance during solar storms. 

While several works~\cite{leo-perf-1,leo-perf-2,leo-perf-3,leo-perf-4,leo-perf-5} have evaluated LEO satellite performance under normal operational conditions, the analysis under solar storms is limited. Recent work~\cite{leonet24-solarstorm} explores the impact of the May 2024 storm on the Starlink network, highlighting the immediate impact on loss and delayed but sustained impact on RTTs. Furthermore, CosmicDance~\cite{cosmicDance} uses real-world measurements to investigate the impact of solar storms on satellite orbits. However, both works stop short of linking orbital dynamics during solar storms with real-time network performance degradation.

Atmospheric drag~\cite{drag-collision} plays a critical role in satellite behavior, particularly during periods of heightened solar activity. Increased solar radiation heats the Earth's upper atmosphere, causing it to expand and increase in density at higher altitudes. This denser atmosphere results in greater drag on LEO satellites, which in turn leads to orbital decay, often observed as a loss of altitude. Satellites must frequently compensate for this decay through propulsion maneuvers to maintain operational altitudes and prevent premature re-entry.

\section{Results}
In this section, we present a detailed analysis of the impact of solar storms on LEO network performance, using fine-grained network measurements and satellite position data. We analyze the regional effects of solar storms on network performance and examine patterns among the most impacted satellites and orbits in the LEO constellation.

\subsection{Experimental Setup}

Our evaluation combines three complementary data sources to quantify and correlate the impacts of solar storms on Starlink's network performance and orbital dynamics. 

\begin{figure}
    \centering
    \includegraphics[width=\linewidth]{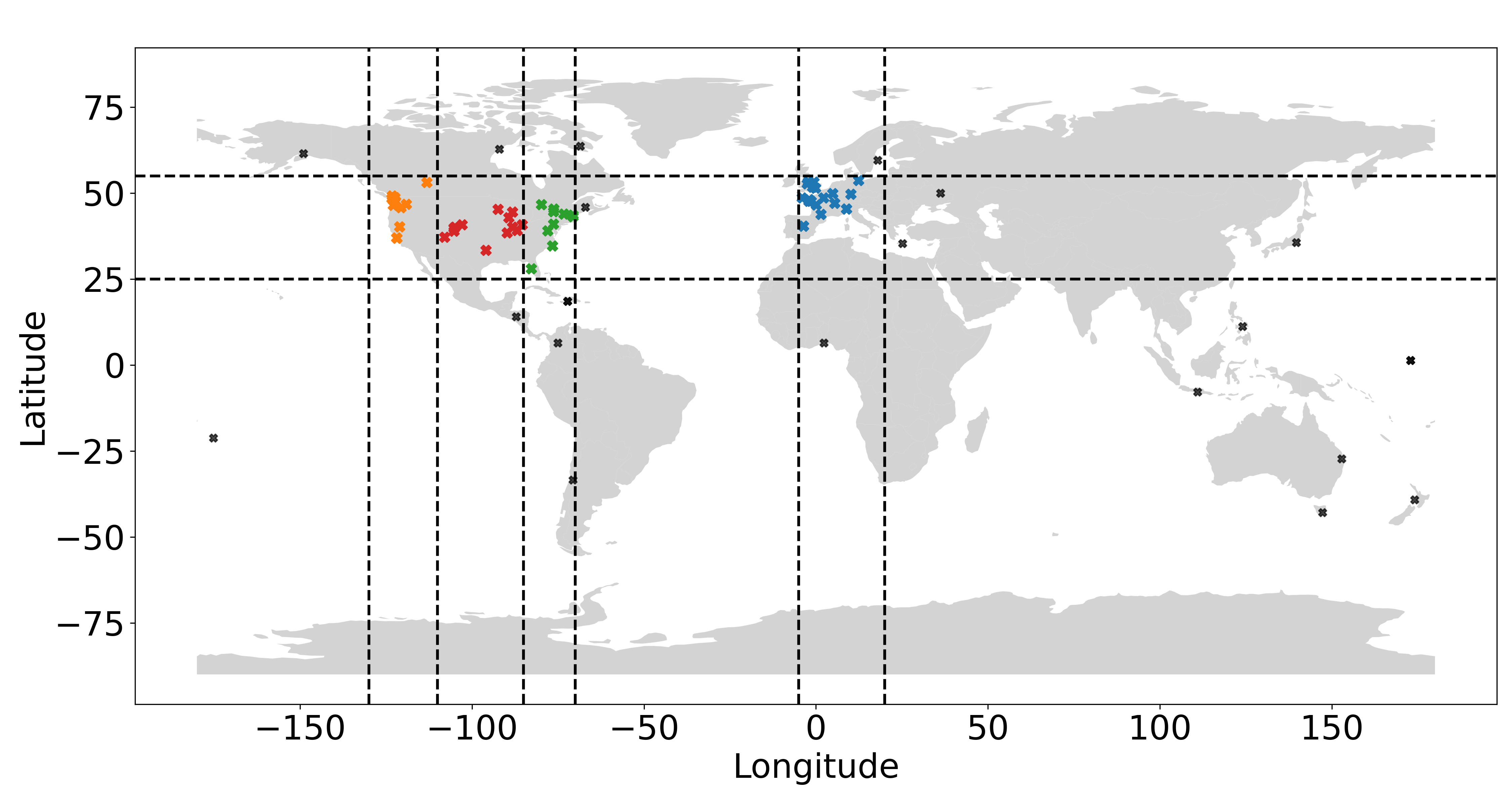}
    \caption{\small The locations of RIPE Atlas probes used in our analysis. The 76 probes  span 23 countries and are connected to the Starlink AS (AS 14593). For regional analysis, we focus on probes in four regions: Europe (blue), US East (green), US Central (red), and US  West (orange).}
    \label{fig:regional_probes}
\end{figure}

\parab{Network Performance Measurement:} We leverage 76 RIPE Atlas probes~\cite{RIPEprobes} connected to the Starlink autonomous system (AS 14593) to capture effects in the network layer. These probes conduct continuous ping measurements every four minutes to two target sets: all DNS root servers and seven terrestrial anchor probes located in Germany, Singapore, the US, and the Netherlands. Figure~\ref{fig:regional_probes} shows the geographic distribution of active Starlink probes during the evaluation period.

\begin{figure}
    \centering
    \includegraphics[width=\linewidth]{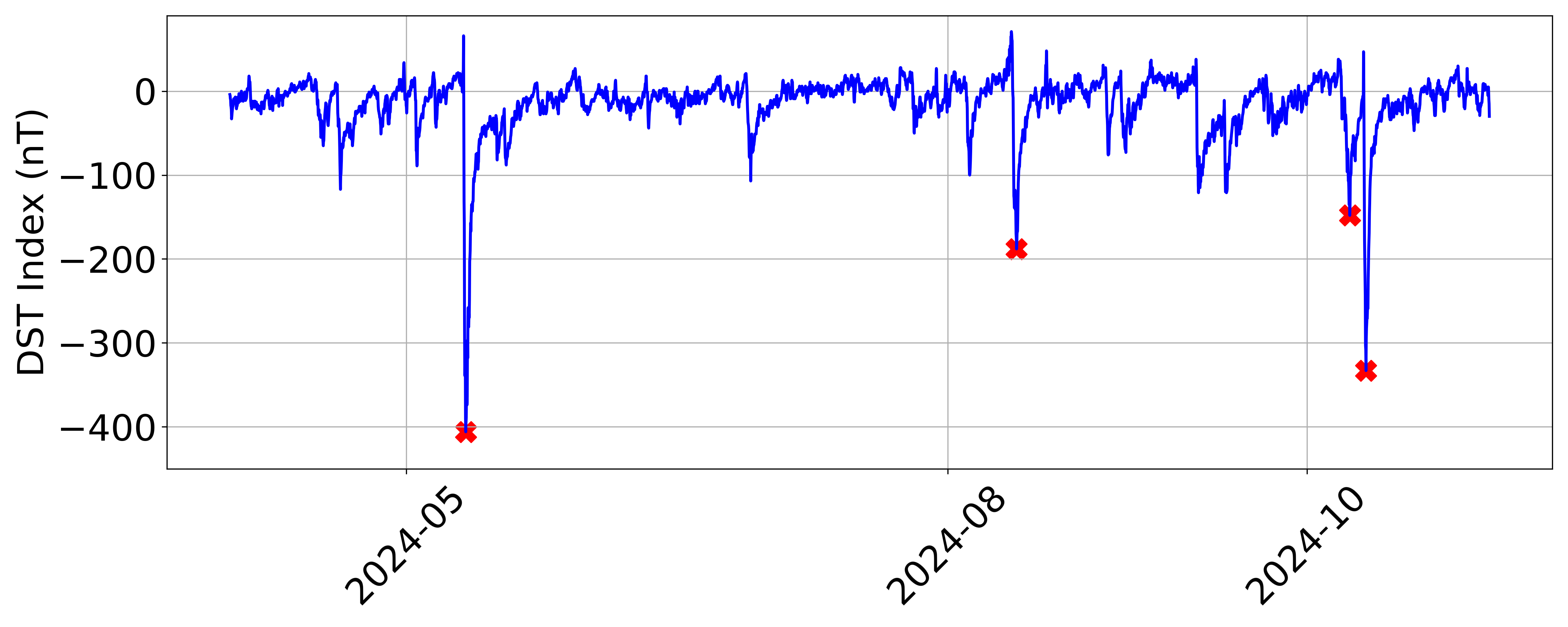}
    \caption{\small Geomagnetic activity in 2024 measured by Dst index. Red markers indicate the four major storms analyzed: May 11\textsuperscript{th} (-406 nT), August 12\textsuperscript{th} (-188 nT), October 8\textsuperscript{th} (-148 nT), and October 11\textsuperscript{th} (-333 nT).}
    \label{fig:dst_index}
\end{figure}

\parab{Solar Storms:} The Disturbance storm time index (Dst)~\cite{kyoto-dst} captures the intensity of geomagnetic disturbance, where more negative values indicate stronger storms. We focus on four major CME events from 2024 (Figure~\ref{fig:dst_index}), selected based on their Dst index: May 11\textsuperscript{th} superstorm (peak Dst of -406 nT at 02:00 UTC), August 12\textsuperscript{th} storm (peak Dst of -188 nT at 16:00 UTC), October 8\textsuperscript{th} storm (peak Dst of -148 nT at 07:00 UTC), and October 11\textsuperscript{th} superstorm (peak Dst of -333 nT at 01:00 UTC). 

\parab{Satellite Orbital Analysis:} We use Two-Line Element set (TLE) data from Space-Track~\cite{SpaceTrack} to quantify altitude variations in Starlink satellites during storm periods. The TLE data provide precise orbital parameters, updated approximately every 10 hours on average, enabling the detection of atmospheric drag-induced altitude changes.

\subsection{Regional Impact Analysis}

\begin{figure*}
    \begin{subfigure}{0.48\textwidth}
        \includegraphics[width=\textwidth]{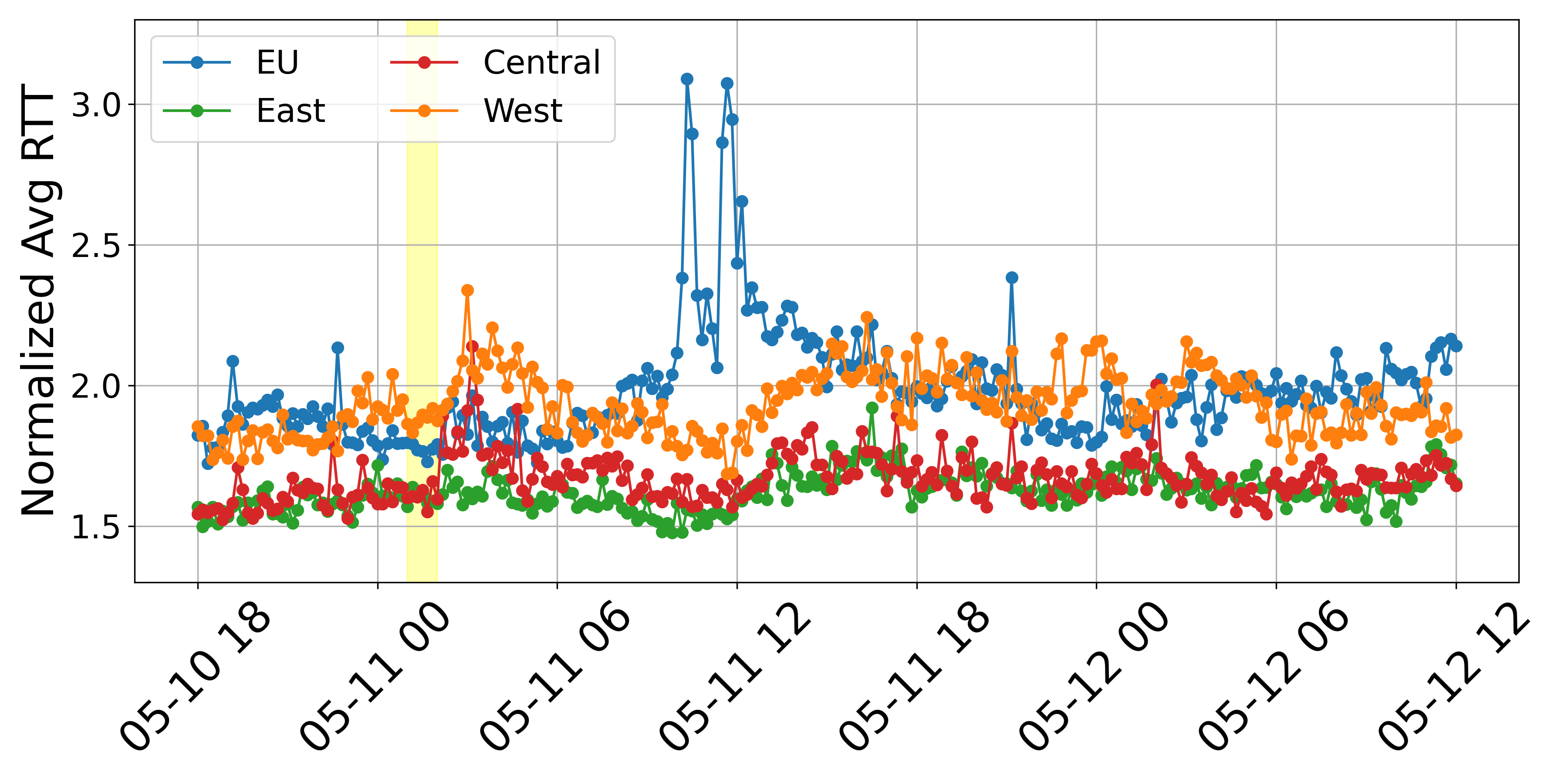}
        \caption[]%
        {{\small May 11\textsuperscript{th} super storm. The date and hour (UTC) are shown on the x-axis.\newline}}
        
        \label{fig:may_11_rtt}
    \end{subfigure}
    \hfill
    \begin{subfigure}{0.48\textwidth}
        \includegraphics[width=\textwidth]{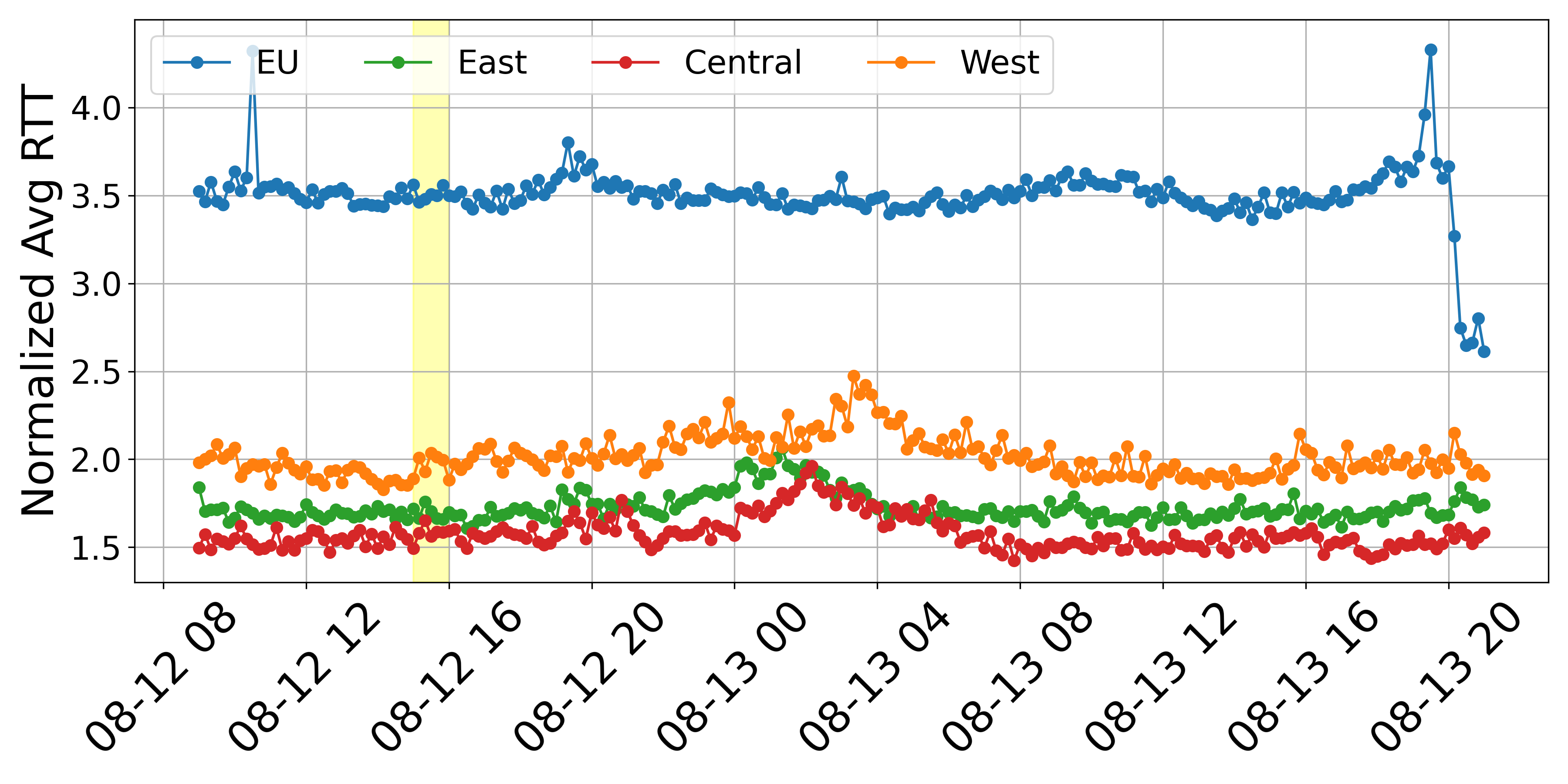}
        \caption[]%
        {{\small August 12\textsuperscript{th} storm. Peak RTTs following the storm occur at 19:00 UTC (Europe), 01:00 UTC (East), 02:00 UTC (Central), and 03:00 UTC (West).}%
        }
        \label{fig:aug_12_rtt}
    \end{subfigure}
    \medskip
    \begin{subfigure}{0.48\textwidth}
        \includegraphics[width=\textwidth]{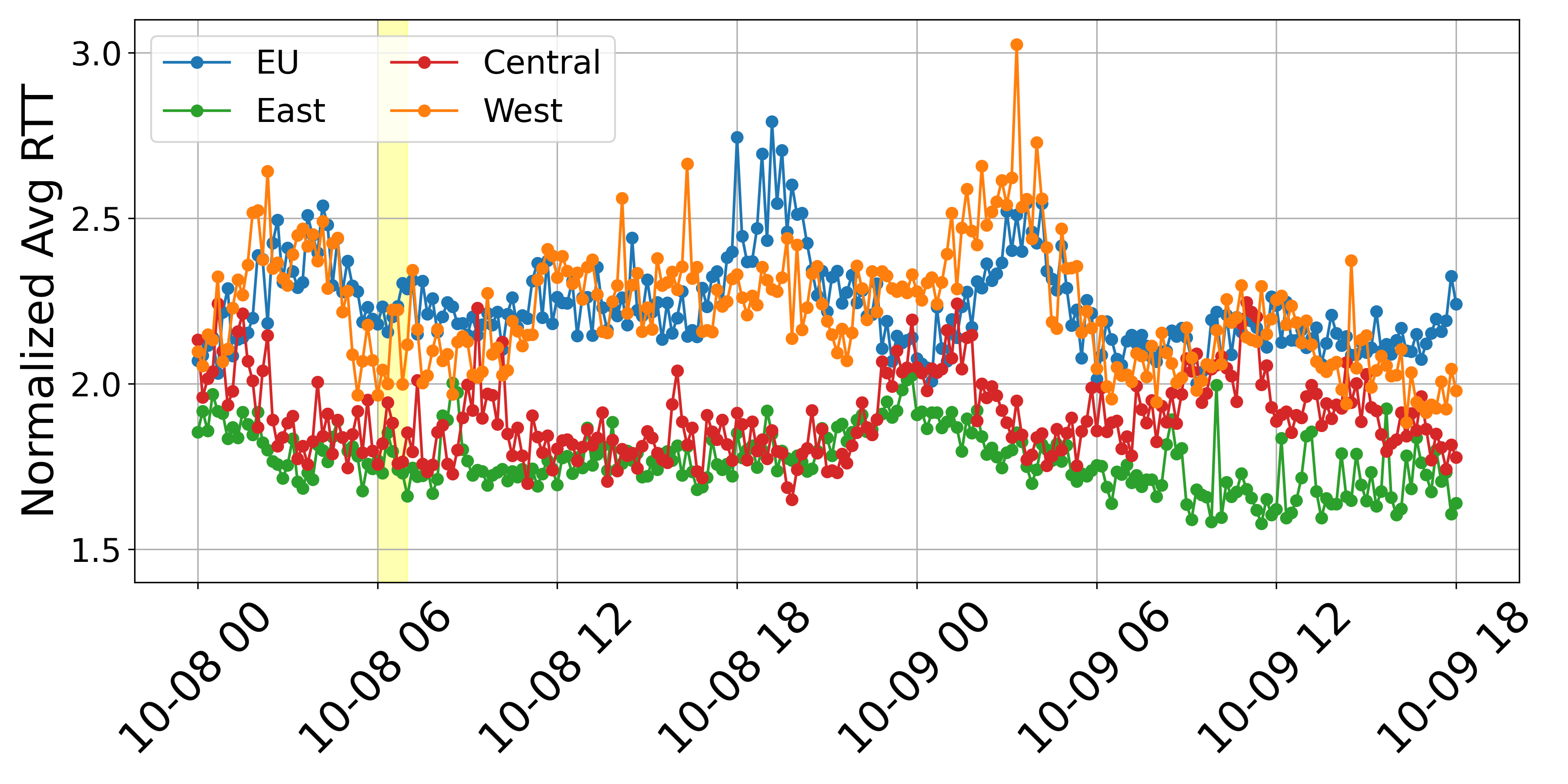}
        \caption[]%
        {{\small October 8\textsuperscript{th} storm. Peak RTTs occur at 18:00-19:00 UTC (Europe), 00:00 UTC (East), 01:00 UTC (Central), and 03:00 UTC (West).}
        }
        \label{fig:oct_8_rtt}
    \end{subfigure}
    \hfill
    \begin{subfigure}{0.48\textwidth}
        \includegraphics[width=\textwidth]{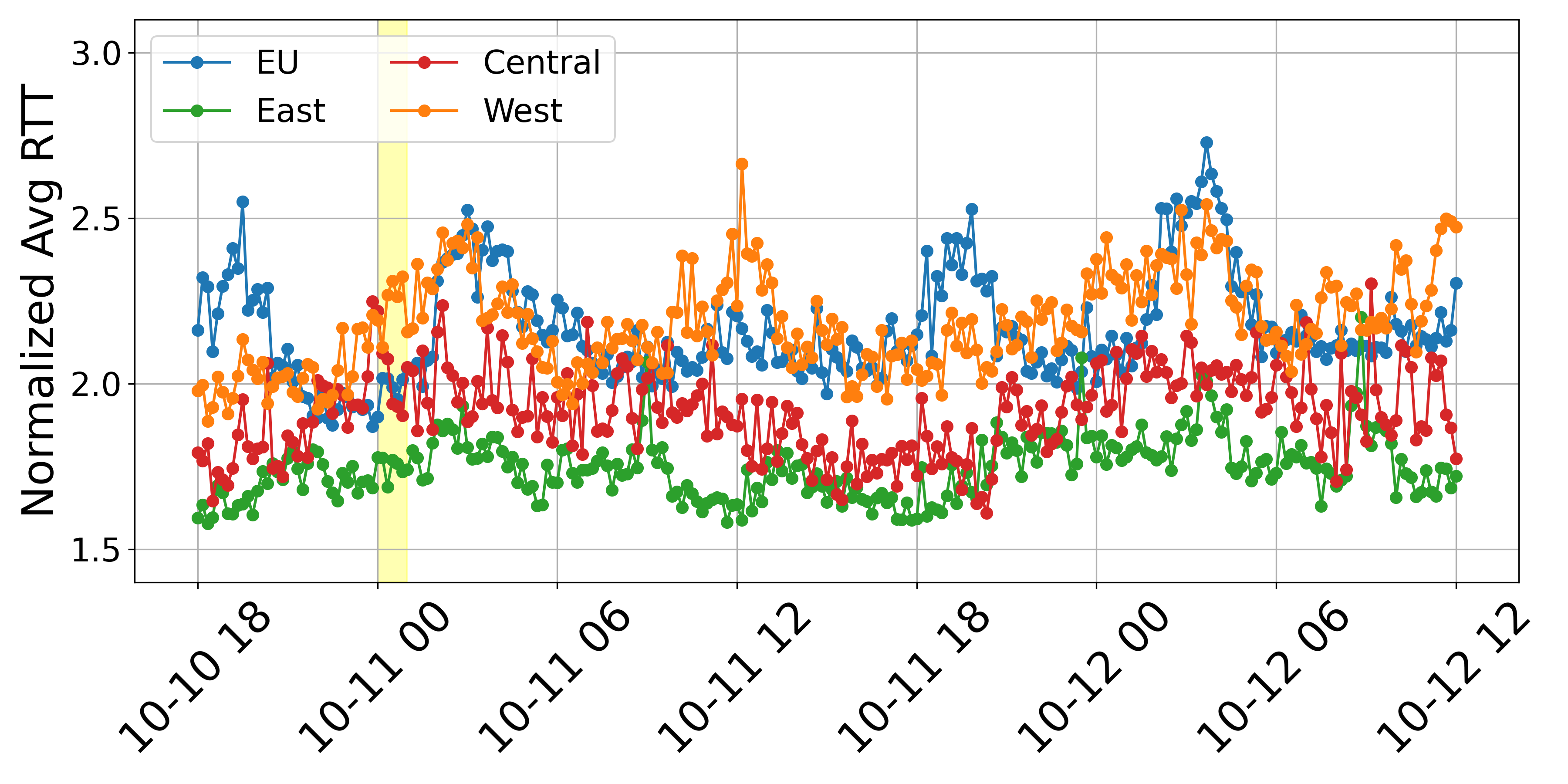}
        \caption[]%
        {{\small October 11\textsuperscript{th} super storm. The effects are less pronounced due to the cumulative effects of the Oct 8\textsuperscript{th} storm.}}
        \label{fig:oct_11_rtt}
    \end{subfigure}
    \caption{\small Normalized RTT across four regions during four geomagnetic storms. Yellow bars indicate 1-hour range storm peaks (peak Dst times). There is a temporal progression in peak impact across regions, aligning with orbital shifts.}
    \label{fig:rtt_plots}
\end{figure*}

We conduct a fine-grained evaluation of the correlation between a probe's geographic location and the severity of impact. We focus on RTT analysis since packet loss manifests as brief spikes that complicate regional comparison.

\parab{Regional Grouping of Probes:} We partition probes into four regions based on geographic concentration: US West (130°W to 110°W longitude), US Central (100°W to 85°W), US East (85°W to 70°W), and Europe (5°W to 20°E). All regions lie in the latitude range between 25°N and 55°N. Figure~\ref{fig:regional_probes} shows the probe distribution across these regions. We exclude probes from the other regions due to small sample sizes.

\parab{Methodology:} For each storm, we calculate the normalized average RTT by dividing the observed RTT by the minimum RTT recorded for each pair of (measurement, probe) during the analysis time frame of 7 days (3 days before the storm, the day of the storm, and 3 days after the storm). This normalization enables cross-regional comparison. We aggregate measurements within 10-minute intervals to capture temporal dynamics while reducing noise.

\parab{Findings:} Figure~\ref{fig:rtt_plots} presents normalized RTT across the four regions during the four storms. Consistent with previous findings~\cite{leonet24-solarstorm}, we observe a delayed increase in RTT that persists for several hours after the peak of the storm across all regions and storms. We do not find any systematic correlation between probe location and impact magnitude. However, we discover a striking temporal pattern in the peak values of RTT across regions. For instance, consider the October 8\textsuperscript{th} storm (Figure~\ref{fig:oct_8_rtt}), peak RTTs occur at 18:00-19:00 UTC in Europe, followed by 00:00, 01:00, and 03:00 UTC in US East, US Central, and US West, respectively.  

This temporal shift can be partially attributed to Starlink's orbital dynamics. Satellite orbits that are directly facing the storm over Europe subsequently pass over US regions as Earth rotates. While a satellite takes around 90 minutes to complete one orbit, the orbit's position will shift approximately 22.5$\degree$ west relative to Earth during the same interval due to Earth's rotation.
As a result, the orbital group above Europe will move over the US regions 5-8 hours later. This might partially explain the westward progression of the RTT peaks, although this interpretation is speculative. 

\begin{figure*}
    \centering
    \begin{subfigure}{0.33\textwidth}
        \includegraphics[width=\textwidth]{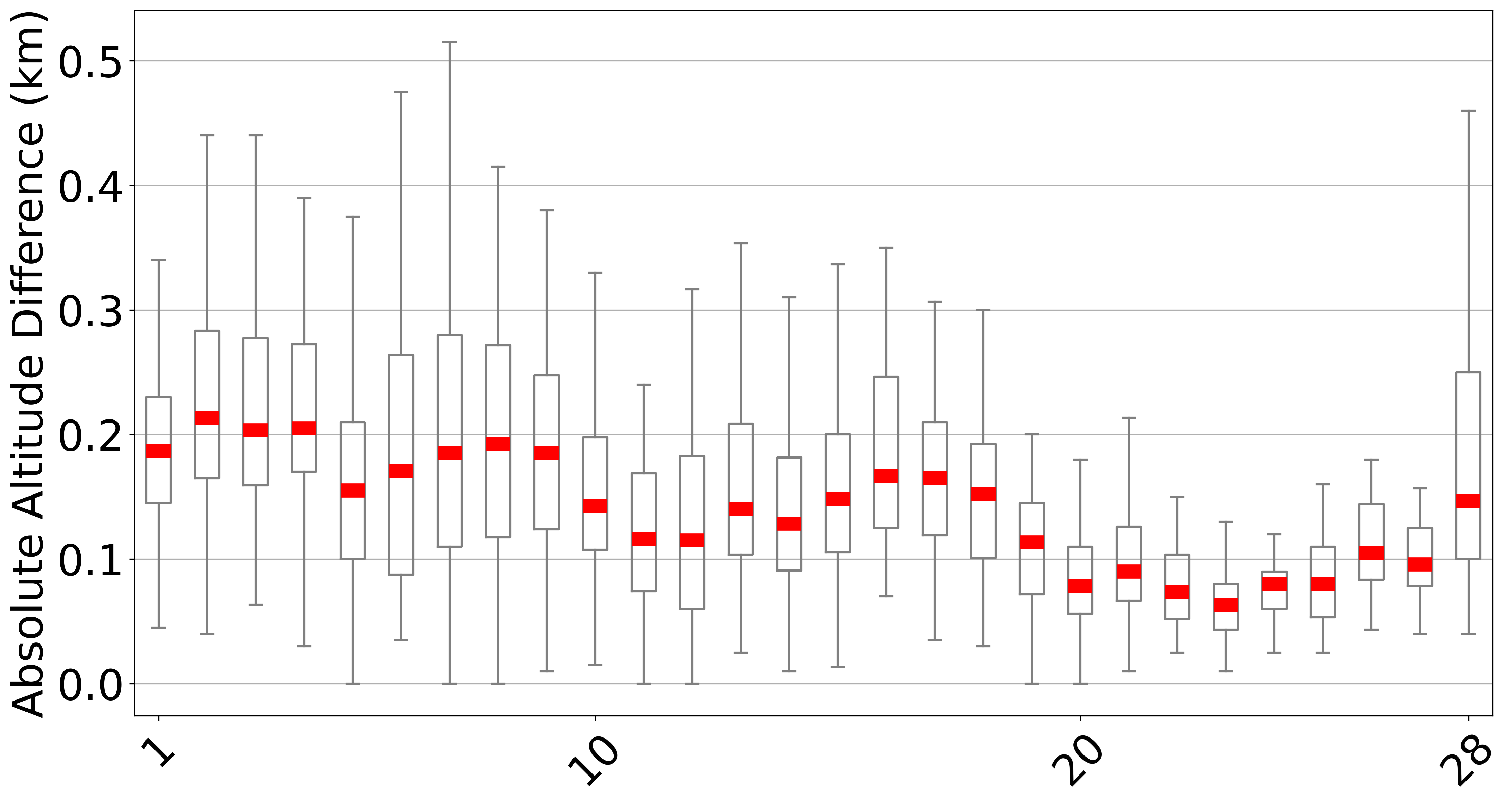}
        \caption[]%
        {{\small 43$\degree$ Inclination Orbits}%
        }
        \label{fig:43_inclination}
    \end{subfigure}
    \medskip
    \begin{subfigure}{0.33\textwidth}
        \includegraphics[width=\textwidth]{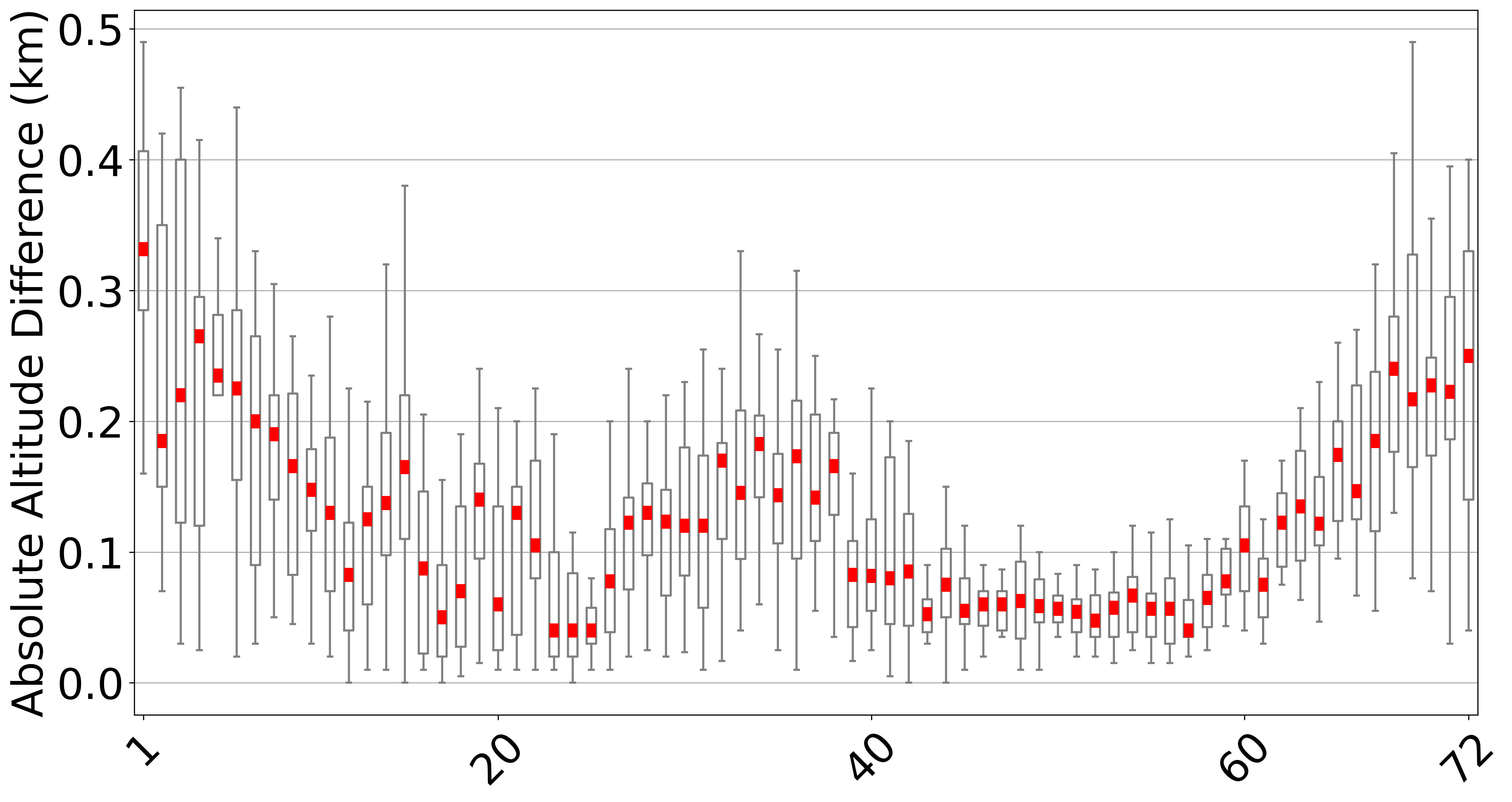}
        \caption[]%
        {{\small 53$\degree$ Inclination Orbits}%
        }
        \label{fig:53_inclination}
    \end{subfigure}
    \medskip
    \begin{subfigure}{0.33\textwidth}
        \includegraphics[width=\textwidth]{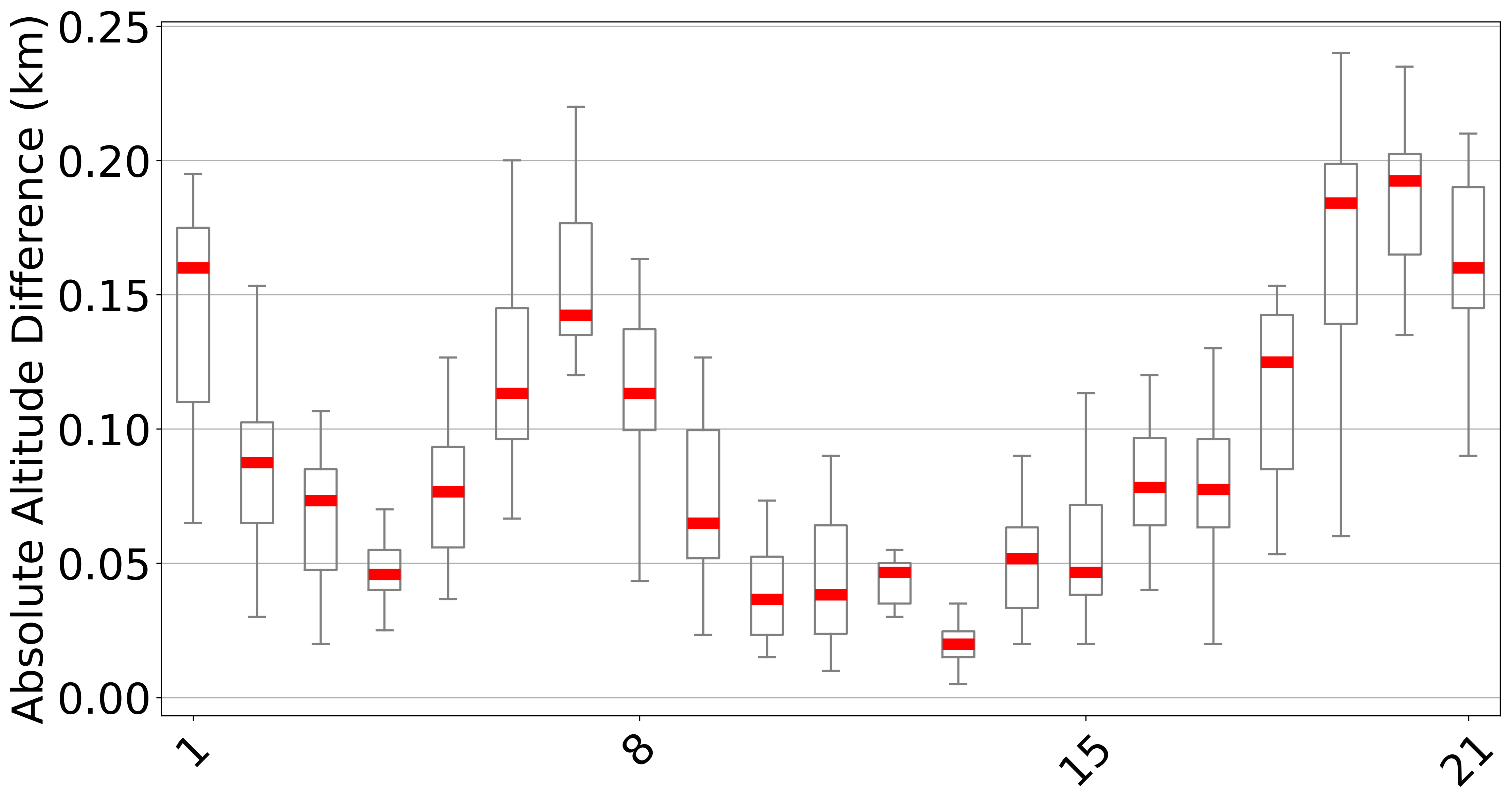}
        \caption[]%
        {{\small 70$\degree$ Inclination Orbits}%
        }
        \label{fig:70_inclination}
    \end{subfigure}
    \caption{\small Altitude changes across orbital groups at three inclinations (43°, 53°, and 70°) during the May 2024 superstorm peak. Box plots show significant variation within inclinations and distinctive "W" patterns across orbits, indicating a systematic impact based on orbit orientation relative to solar storm direction. Higher-resolution images for four orbital inclinations (43°, 53°, 70°, and 97.6°) during the storm period are provided in Figure~\ref{fig:inclinations_plot_appendix}, with corresponding plots from a non-storm period included for comparison in Figure~\ref{fig:inclinations_plot_baseline_ylim_appendix} in the Appendix.}
    \label{fig:inclinations_plot}
\end{figure*}

\begin{figure}
    \centering
    \begin{subfigure}{0.23\textwidth}
        \includegraphics[width=\textwidth]{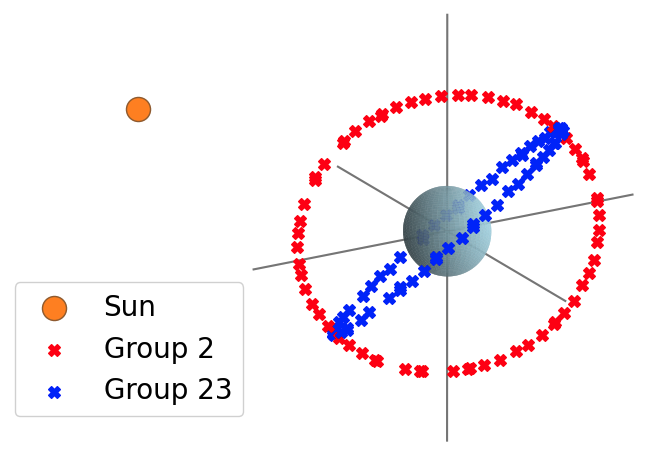}
        \caption[]%
        {{\small 90$\degree$ Phase Difference}%
        }
        \label{fig:phase_90}
    \end{subfigure}
    \hfill
    \begin{subfigure}{0.23\textwidth}
        \includegraphics[width=\textwidth]{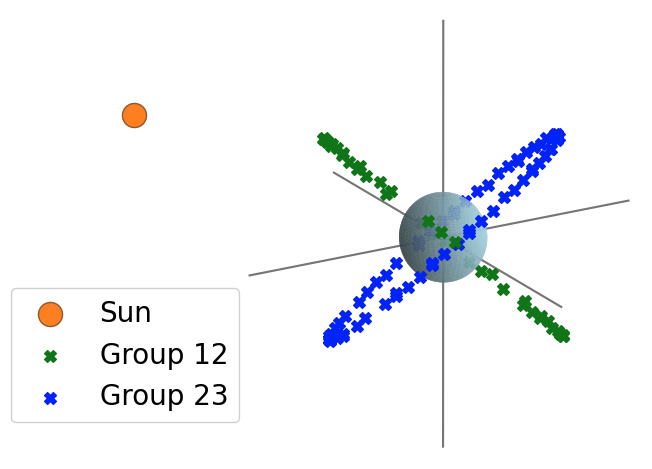}
        \caption[]%
        {{\small 180$\degree$ Phase Difference}%
        }
        \label{fig:phase_180}
    \end{subfigure}
    \caption{\small Orbit positions at 43$\degree$ inclination during May 2024 superstorm. (a) The most (group 2) and the least impacted orbits (group 23) have a phase difference of 90$\degree$. (b) Group 12 and group 23 have $\approx$180$\degree$ of phase difference. Note that the sizes of the Sun, the Earth, and the orbits are not to scale, but their relative orientations are accurate.}
    \label{fig:inclination_most_least}
\end{figure}

Note that in the plots, some normalized RTT values are consistently above 1.0 since the normalization is with respect to the minimum across 7 days, while the visualization is over 2 days. Hence, the minimum may exist beyond the visualized interval. In addition, elevated values in certain regions reflect cumulative effects from preceding storms, since geomagnetic superstorms typically occur in clusters. The larger variance on October 11\textsuperscript{th} is likely due to the cumulative effect from the preceding storm on October 8\textsuperscript{th}.

\subsection{Inter-Orbital Impact Analysis}

Inspired by the time shift observed in peak network performance degradation across regions, we next investigate the orbit-level impact. We analyze altitude shifts of satellites per orbit using Two-Line Element (TLE) data, calculating the net magnitude of altitude changes in two consecutive TLEs for all satellites within each orbit. These altitude shifts reflect variations in atmospheric drag that correlate with network performance degradation.

\parab{Data processing:} We leverage CosmicDance~\cite{cosmicDance} to calculate the altitude of satellites from TLE data. There were $\approx$5800 Starlink satellites with TLE data during May 2024. However, to focus on the altitude change caused by the solar storm, we remove the satellites in the orbit-raising phase and those not in the planned orbital altitude, and analyze only the remaining 4708 satellites. To derive the position of the satellite at a specific time when TLE is not available, we use the SGP4 orbit propagator~\cite{spg4} to propagate the satellite position from the most recent TLE ephemeris. We group satellites by inclination, altitude, and right ascension of the ascending node (RAAN) to identify those having the same orbital path.

\parab{Patterns in Orbital Impact:} Figure~\ref{fig:inclinations_plot} presents box plots of altitude changes of the satellite across orbits at three inclinations (43°, 53°, and 70°) during the May 2024 superstorm peak (May 11\textsuperscript{th}). Detailed plots including the fourth inclination of  97.6° (Figure~\ref{fig:inclinations_plot_appendix}) and corresponding plots from a non-storm period for comparison  (Figure~\ref{fig:inclinations_plot_baseline_ylim_appendix}) are given in the Appendix. We calculate the change in altitude between consecutive TLE data on May 11\textsuperscript{th}, averaging the values if multiple TLEs are available for the day. This altitude difference reflects both solar storm-induced decay and human-controlled orbital adjustments. We observe significant variations in the severity of impact across orbits within the same inclination, with certain orbits experiencing substantially higher altitude changes than others.

The data reveals a distinctive ``W'' pattern in the box plots across all inclinations, indicating systematic variation in orbital impact based on orbit positioning. Orbits with the highest altitude changes contain satellites positioned to be closer to the Sun, experiencing solar radiation flux directly during the storm. Conversely, orbits with minimal impact had more satellites positioned farther away from the Sun.

\parab{Phase Relationship Analysis:} To validate the hypothesis that orbital orientation with respect to the sun influences the extent of impact, we examine phase relationships between high-impacted and low-impacted orbits during the peak of the May 2024 superstorm. Note that the longitude-based phase difference between adjacent orbits depends on the number of orbits at each inclination. For example, at 53$\degree$, there are 72 orbits, leading to a phase shift between adjacent orbits of $360/72 = 5\degree$. 

In Figure~\ref{fig:inclinations_plot}, analyzing the mean absolute altitude change across various orbits at each inclination, we find consistent phase shifts of 80 - 100$\degree$ or 260 - 280$\degree$ (equivalent to -80$\degree$ to -100$\degree$) between orbits experiencing maximum and minimum altitude changes. For instance, orbit group 2 and group 23 at 43$\degree$ inclination, with the maximum altitude change and minimum altitude change, respectively, exhibit a 90$\degree$ phase shift. Orbit group 1 and group 52, with maximum and minimum altitude change, respectively, at a 53$\degree$ inclination, have around a 260$\degree$ phase shift, which is equivalent to -100$\degree$. This $\approx$90$\degree$ phase separation between the most and least impacted orbits across all inclinations confirms that orbital impact severity correlates with orbital orientation relative to incoming solar radiation.

Figure~\ref{fig:phase_90} shows positions for the most and least impacted orbits at 43$\degree$ inclination during the May 2024 superstorm. The most affected group 2 has more satellites facing the sun, while the least affected group 23 is oriented $\approx$90$\degree$ from it. Additionally, in Figure \ref{fig:inclinations_plot}, we observe variations in the extent of dips on the two halves of the ``W'' pattern. This can also be explained based on the orientation of the orbital plane with respect to the Sun. Figure~\ref{fig:phase_180} shows that orbit group 12, with $\approx$180$\degree$ phase difference with respect to group 23, has nearly half of the satellites oriented toward the Sun. Hence, the dip around group 12 is less pronounced than that around group 23 in the ``W'' pattern. Thus, orbital analysis provides evidence that solar storm impacts on LEO satellites depend critically on satellite orientation relative to the solar flux direction.

\subsection{Geospatial Impact Analysis}

While our previous analysis demonstrates differences across orbital groups, we next delve deeper to examine the geolocation of the most affected satellites during the storm. We again focus on the May 2024 superstorm. To identify satellites that experienced significant altitude changes at each inclination, we establish our impact threshold as the 95\textsuperscript{th} percentile of average altitude changes observed on the day of the peak storm (May 11\textsuperscript{th}). We identify the altitude change threshold for the 95\textsuperscript{th} percentile as 0.325 km, the 97\textsuperscript{th} percentile as 0.367 km, and the 99\textsuperscript{th} percentile as 0.45 km.

\begin{figure}
    \centering
    \includegraphics[width=\linewidth]{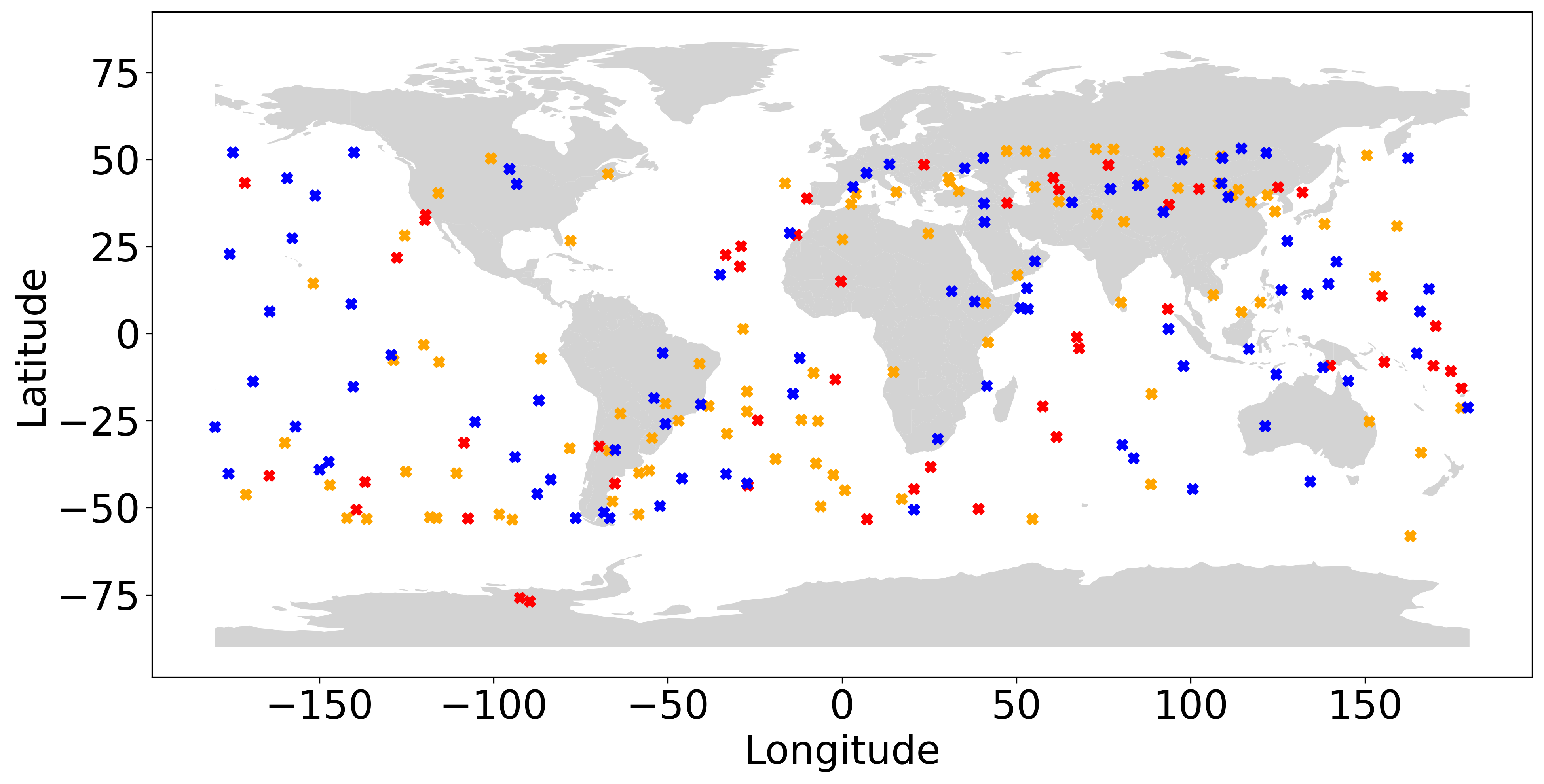}
    \caption{\small Spatial distribution of most impacted satellites at peak Dst during May 2024 superstorm. Impact severity shown by color: red (>99\textsuperscript{th} percentile), orange (97-99\textsuperscript{th} percentile), blue (95-97\textsuperscript{th} percentile).}
    \label{fig:impacted_satellites_may}
\end{figure}

\begin{figure}
    \begin{subfigure}{0.23\textwidth}
        \includegraphics[width=\textwidth]{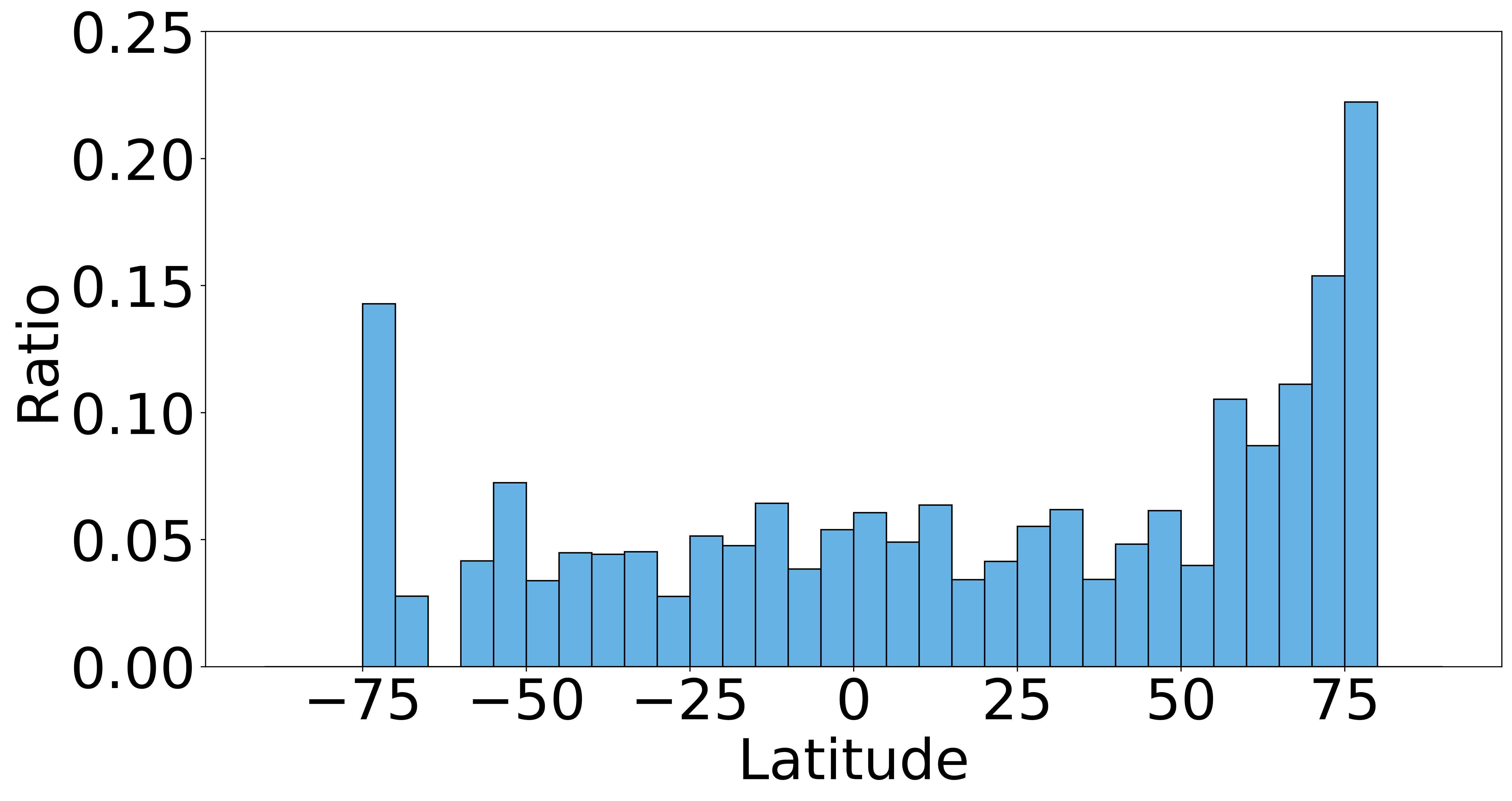}
        \caption[]%
        {{\small Latitude}%
        }
        \label{fig:aug_12_lat}
    \end{subfigure}
    \hfill
    \begin{subfigure}{0.23\textwidth}
        \includegraphics[width=\textwidth]{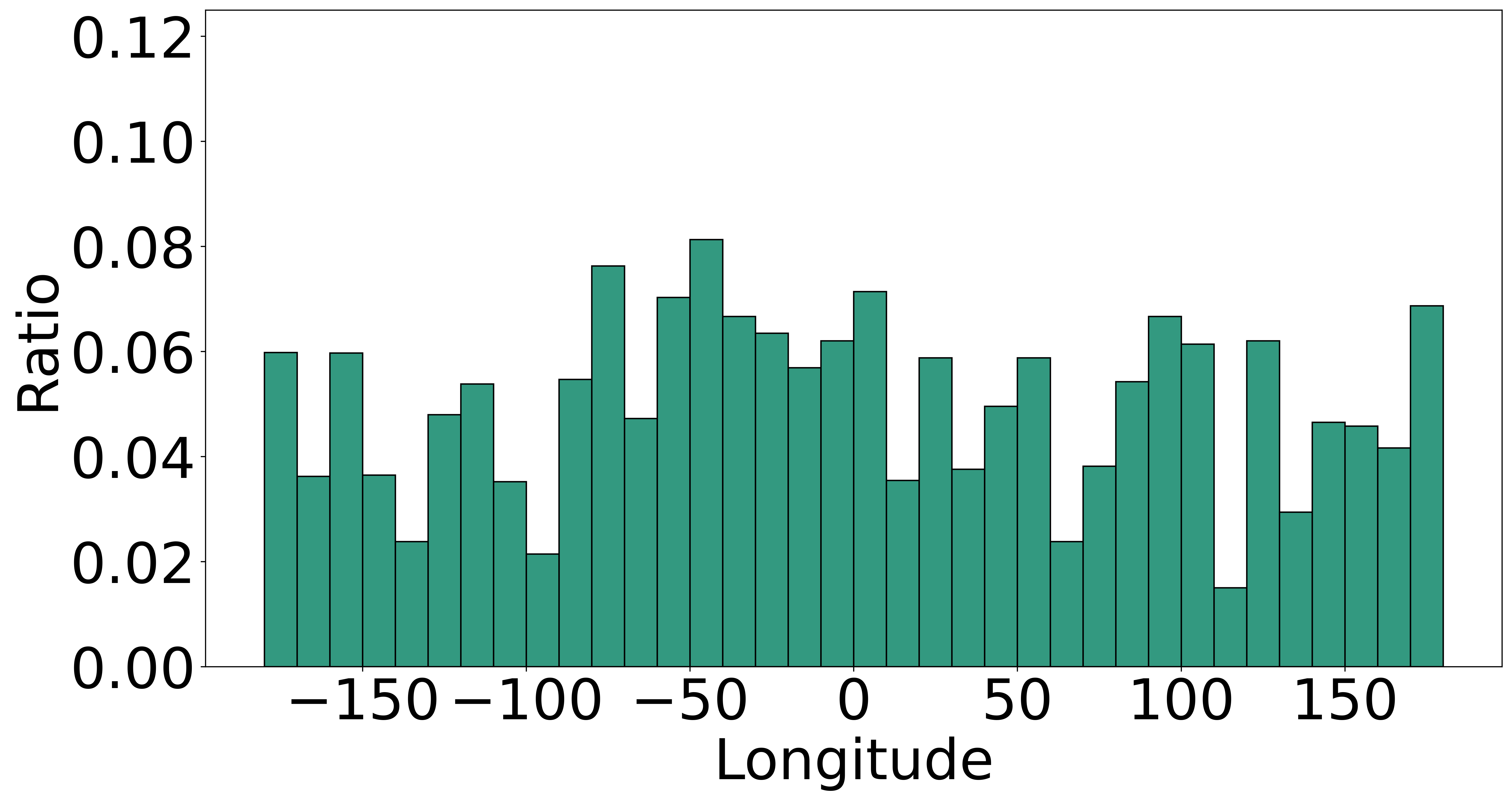}
        \caption[]%
        {{\small Longitude}%
        }
        \label{fig:aug_12_lon}
    \end{subfigure}
    \caption{\small The fraction of highly impacted satellites across latitudes and longitudes during the August 2024 event. We observe a higher impact probability at higher latitudes. Furthermore, there is a higher impact on the region experiencing daytime during peak Dst. }
    \label{fig:lat_lon_plots}
\end{figure}

Figure~\ref{fig:impacted_satellites_may} shows the positions of the most severely impacted satellites across all inclinations on May 11\textsuperscript{th} at approximately 02:00 UTC, coinciding with the peak Dst index. 

\parab{Spatial Impact Patterns:} Our analysis reveals interesting patterns in the spatial distribution of impacted satellites, which can be broadly classified into three distinct groups. First, most of the severely impacted satellites are positioned at higher latitudes, consistent with the enhanced atmospheric drag in polar regions that occurs during geomagnetic storms. Second, we observe a significant impact over the South Atlantic region, where the South Atlantic Anomaly (SAA)~\cite{saa,saa-2} is known to create enhanced exposure due to a weakened Earth's magnetic field. The third subset of highly impacted satellites is primarily located over regions that experienced daytime during the storm's peak Dst at 2:00 am UTC (notably Asia and the Pacific), suggesting enhanced atmospheric drag on the sun-facing side of Earth.

To validate our hypothesis regarding the categories, we further analyze the August 12\textsuperscript{th} 2024 storm. This secondary case study demonstrates the generalizability of our findings across different storm events and examines patterns during a storm with relatively lower Dst magnitude compared to the May 2024 superstorm. Figures~\ref{fig:aug_12_lat} and \ref{fig:aug_12_lon} present histogram analyses showing the fraction of highly impacted satellites among total satellites across latitudes and longitudes for the August 2024 event. Again, we observe that satellites at higher latitudes have significantly higher probabilities of being impacted by the storm. Additionally, it reveals a higher distribution during the daytime at the storm's peak Dst, at 16:00 UTC, between 0$\degree$ and -50$\degree$ longitudes.

\begin{figure}
    \centering
    \includegraphics[width=\linewidth]{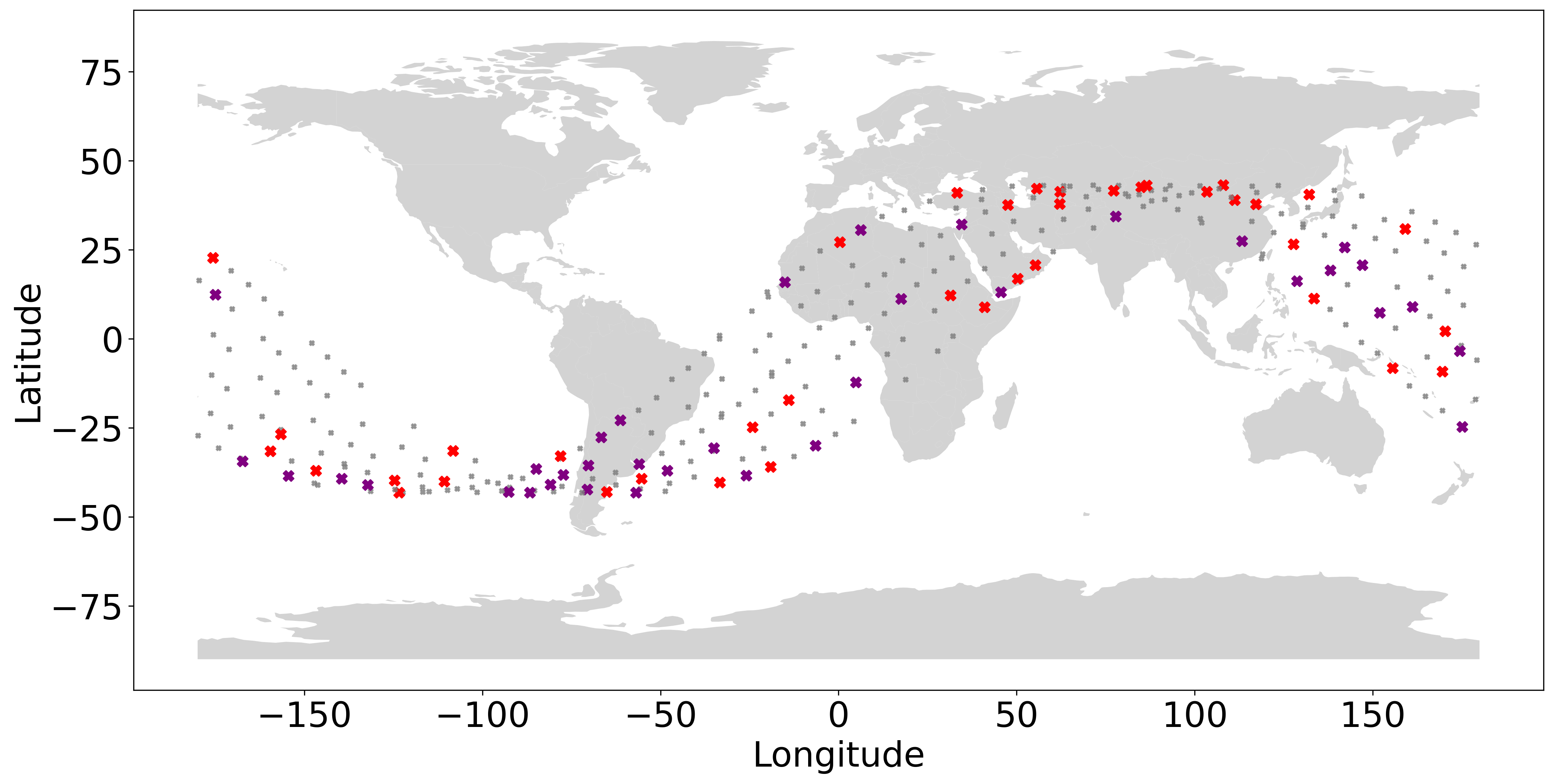}
    \caption{\small Position of satellites in groups 2 to 7 at inclination 43$\degree$ at the Dst peak of May 2024 superstorm. Highly impacted satellites on May 11\textsuperscript{th} marked in red, and high-impacted satellites on May 12\textsuperscript{th} marked in purple.}
    \label{fig:two_day_diff}
    \vspace{-2mm}
\end{figure}

\parab{Clustered Pattern Across Multiple Days: }
While Figure~\ref{fig:inclinations_plot} relies on TLE data from May 11\textsuperscript{th}, a similar pattern is observed on May 12\textsuperscript{th}, but with a different set of highly impacted satellites. Among the 236 satellites that are over the 95\textsuperscript{th} percentile of altitude change on May 11\textsuperscript{th}, only 30 satellites were also present in the highly impacted subset on May 12\textsuperscript{th}. Interestingly, Figure~\ref{fig:two_day_diff} shows that the most impacted satellites across the two days, even when they are distinct, are positioned near each other. Thus, we observe that significant altitude changes propagate within and across orbits, affecting the neighboring satellites. We attribute this to the self-driving nature of LEO constellations, which perform automatic orbital maneuvers to calibrate satellite orbits and avoid collisions~\cite{self-driving-LEO}.

\section{Discussion}

In this section, we discuss the implications of our findings on network design and resilience of LEO constellations.

\parab{Leveraging Predictability of Impact: } Our findings demonstrate that satellites at high latitudes and those with direct solar exposure during storm peaks experience the most significant orbital decay, corroborating previous thermosphere ionosphere coupled simulations \cite{drag-study}. 
This predictability presents critical opportunities for enhancing network resilience through proactive management strategies.

Operators can implement preemptive network adaptations by integrating solar storm forecasts with our identified vulnerability patterns. Networks can dynamically reconfigure topologies, preposition backup communication paths through less-vulnerable satellites, and implement priority-based handover mechanisms that favor satellites likely to be in low-impact orbits. By predicting when and where satellites are likely to experience orbital degradation, networks can dynamically adapt routing protocols and create resilience zones by identifying satellite clusters likely to maintain stable performance during storms.

However, not all satellites traversing high-risk regions experience uniform altitude degradation. This variability is due to highly localized atmospheric density deviations and satellite-specific factors. This heterogeneity necessitates the development of probabilistic impact models that capture the stochastic nature of drag effects, enabling networks to make risk-aware routing decisions that balance performance optimization with reliability guarantees during space weather events.

\parab{The Propagation of Impact across Satellites: } Our analysis reveals that different sets of satellites experience the most severe impacts across consecutive storm days, with only 30 out of 236 highly impacted satellites on May 11\textsuperscript{th} remaining in the high-impact category on May 12\textsuperscript{th}. While atmospheric density pockets could theoretically affect multiple satellites across days, the observed pattern of neighboring satellites being sequentially impacted suggests a different cause. We attribute this clustered propagation primarily to the self-driving nature of LEO constellations. Starlink employs a proprietary onboard autonomous control system to manage continuous orbit maintenance, collision avoidance, and inter-shell maneuvers. However, excessive or unnecessary maneuvers can lead to network topology instability and performance degradation~\cite{self-driving-LEO}. Analysis of TLE data reveals that when satellites experience altitude loss due to increased atmospheric drag, Starlink responds by temporarily raising the affected satellites above their nominal altitude. These satellites typically return to their original altitude within 1–2 days. This corrective action triggers a cascading effect, with orbital adjustments propagating across neighboring satellites in both spatial and temporal dimensions. Full stabilization of the orbit often takes 3–4 days. These dynamic adjustments can disrupt satellite links and routing paths, contributing to performance issues such as a sustained increase in round-trip time (RTT).

This finding raises fundamental questions about the suitability of today's autonomous constellation management during extreme space weather events. The self-driving algorithms, optimized for normal operations, may inadvertently amplify storm impacts by triggering chains of orbital adjustments. Network operators should investigate implementing storm-aware autonomous control modes that temporarily modify or disable certain self-driving behaviors during severe geomagnetic disturbances and introduce collective decision-making algorithms that consider space weather conditions when planning constellation-wide maneuvers. 

\parab{A Real-time Monitoring Framework:} A monitoring framework that compares predicted solar storm impact models against observed network performance metrics can enable the iterative refinement of probabilistic impact models in real time, in turn facilitating network adaptations. For example, such a monitoring system could provide early warning capabilities by detecting initial impacts in eastern regions before they propagate westward. This real-time validation loop would improve the accuracy of the 3-day forecast window typically available for solar storm predictions, allowing operators to make more informed decisions about preemptive network reconfigurations and resource allocation.

\parab{Limitations:}
Our analysis is constrained by the sparsity of available data sources. The satellite TLE data, with an average gap of 10 hours between reported positions, makes it difficult to determine the precise timing of satellite decay. Similarly, latency analysis is limited by the coarse resolution of ping measurements, taken every 4 minutes, which fail to capture short-term variations in loss and latency. Additionally, the sparse distribution of measurement probes at lower latitudes results in reduced data coverage in those regions.

The opaque nature of the Starlink system further constrains our analysis. Although RIPE Atlas measurements provide the geographic coordinates of probes, they do not reveal which satellite is serving a given probe at any point in time. Additionally, the lack of traceroute support within the Starlink network prevents us from identifying the satellite hops involved along a given path.

The observed impact of solar storms on satellite performance is also highly non-uniform. While a higher concentration of affected satellites appears at higher latitudes, not all satellites in these regions are impacted. Even in the most affected regions, no more than 20\% of satellites experienced significant altitude changes. This uneven impact is likely due to local variations in atmospheric conditions along individual satellite paths. These findings underscore the need for a more detailed analysis to gain a deeper understanding of the fine-grained dynamics of satellite behavior under solar storm conditions.

\section{Conclusion}
In this paper, we present a fine-grained analysis of the impact of solar storms on LEO networks by correlating real-world network performance data with satellite orbital dynamics across multiple storm events. Our study revealed significant spatial, temporal, and orbital heterogeneity in how solar activity impacts LEO network performance, highlighting vulnerable satellite regions, including high latitudes, sun-facing orbits, and the South Atlantic Anomaly. Future work should explore the design of adaptive network solutions as well as cross-layer control loops that integrate atmospheric drag forecasts, real-time RTT feedback, and autonomous satellite maneuver planning to build next-generation resilient satellite networks. While this study focuses on Starlink---the largest LEO constellation currently in operation---our methodology can be extended to analyze other systems such as OneWeb~\cite{OneWeb} and Kuiper~\cite{Kuiper}.

\bibliographystyle{ACM-Reference-Format}
\bibliography{references}

\appendix
    
\begin{figure*}
    \centering
    \begin{subfigure}{0.48\textwidth}
        \includegraphics[width=0.9\textwidth]{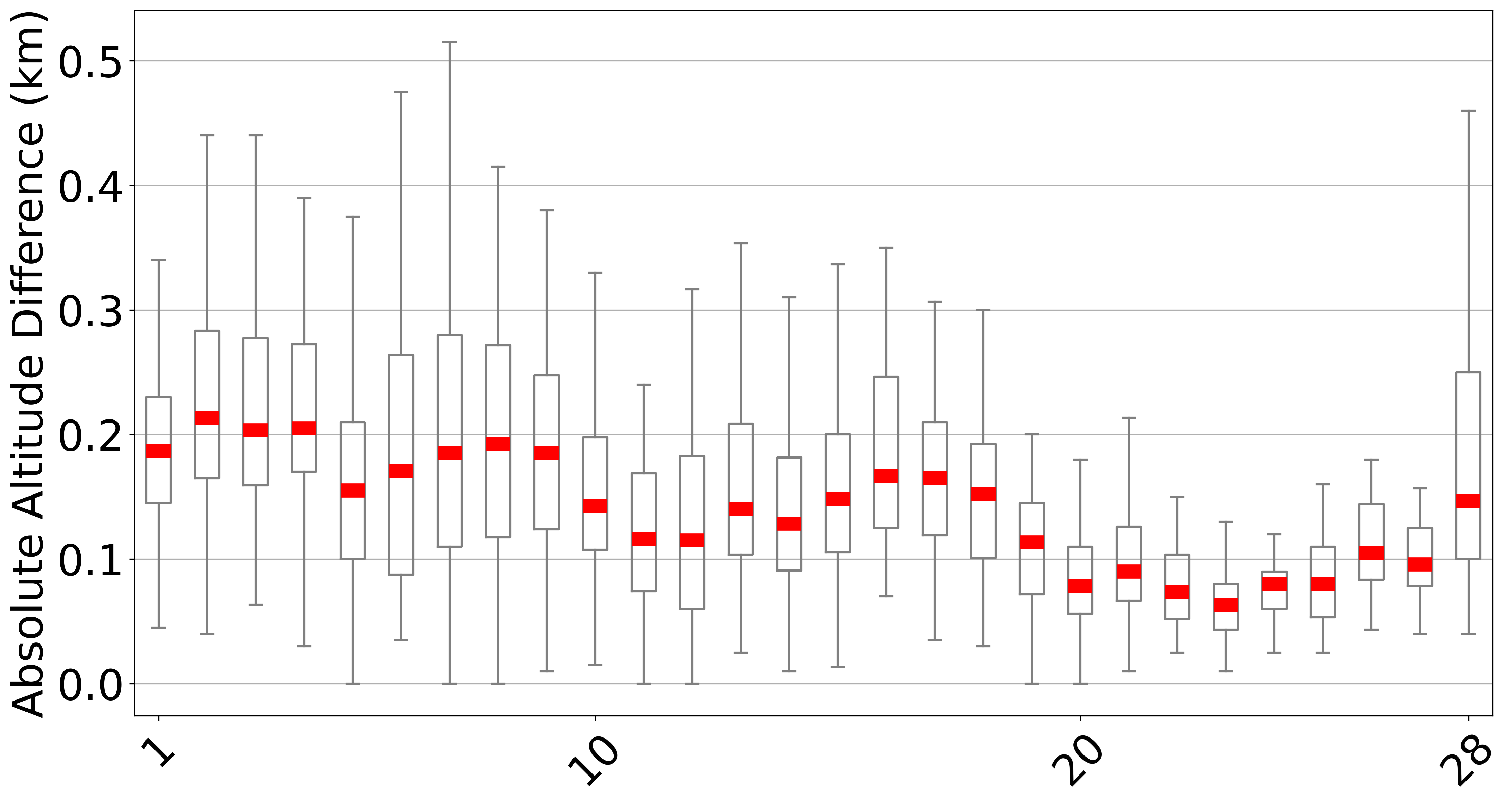}
        \caption[]%
        {{\small 43$\degree$ Inclination Groups}%
        }
        \label{fig:43_inclination_appendix}
    \end{subfigure}
    \hfill
    \begin{subfigure}{0.48\textwidth}
        \includegraphics[width=0.9\textwidth]{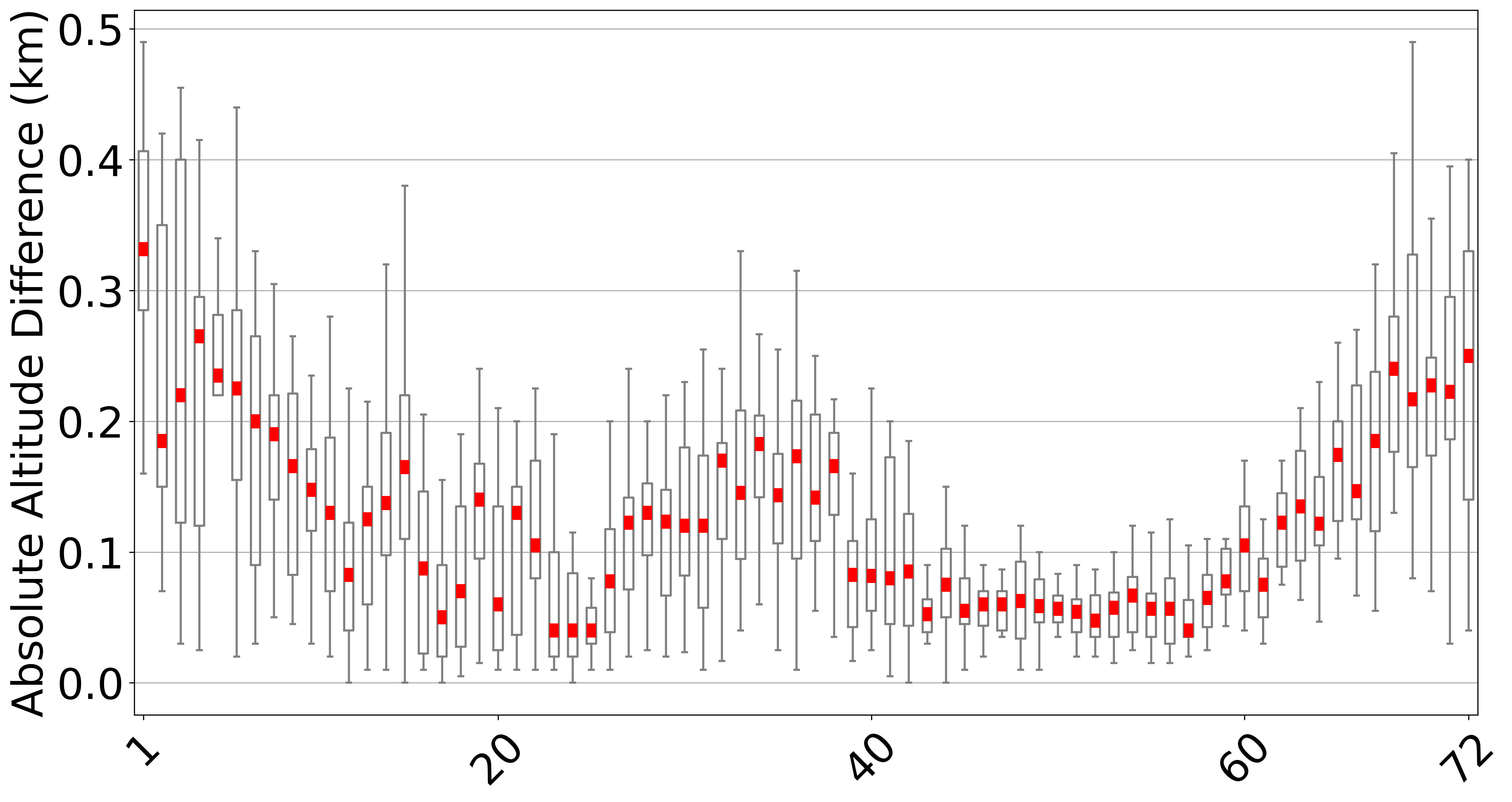}
        \caption[]%
        {{\small 53$\degree$ Inclination Groups}%
        }
        \label{fig:53_inclination_appendix}
    \end{subfigure}
    \medskip
    \begin{subfigure}{0.48\textwidth}
        \includegraphics[width=0.9\textwidth]{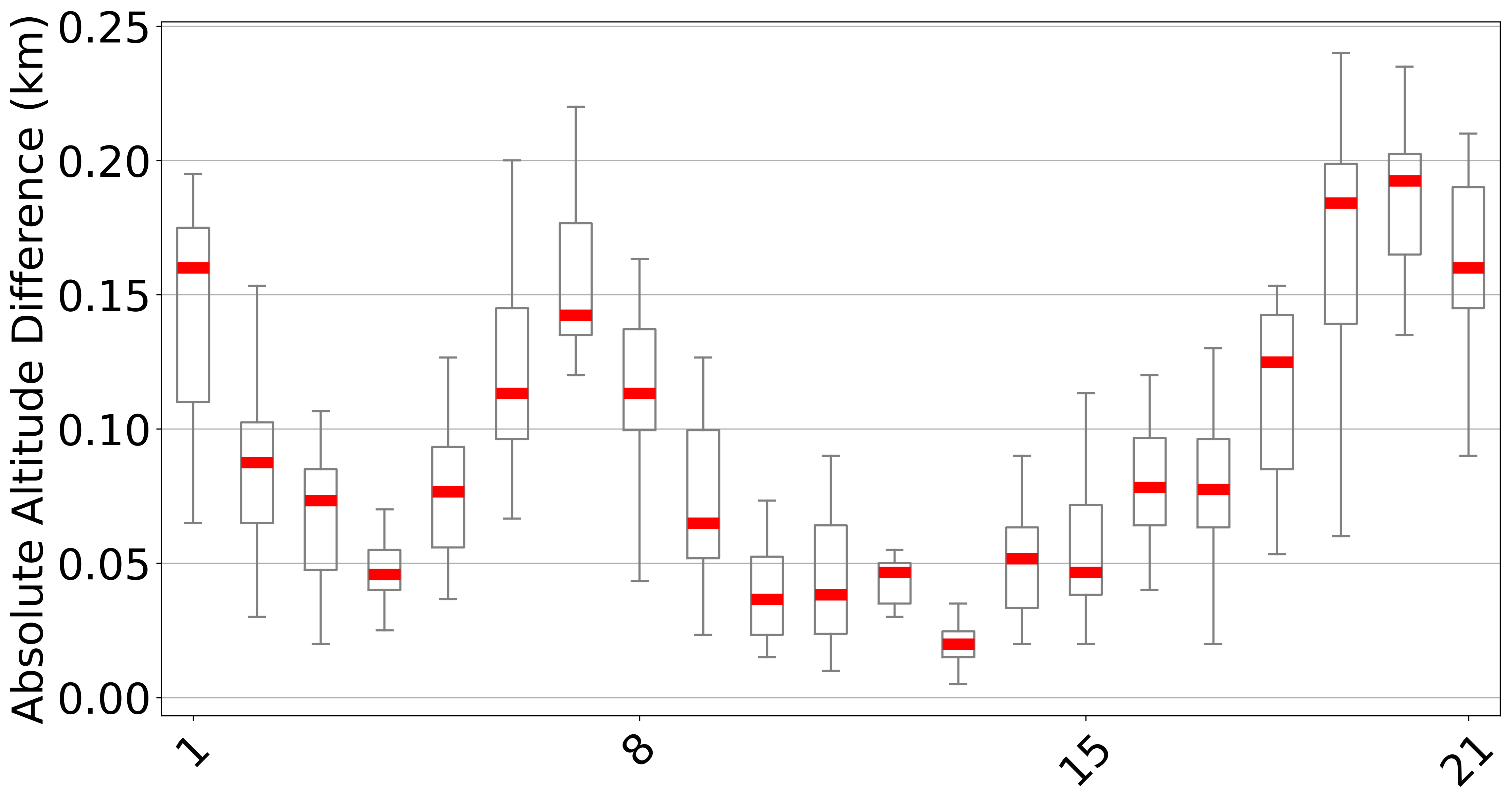}
        \caption[]%
        {{\small 70$\degree$ Inclination Groups}%
        }
        \label{fig:70_inclination_appendix}
    \end{subfigure}
    \hfill
    \begin{subfigure}{0.48\textwidth}
        \includegraphics[width=0.9\textwidth]{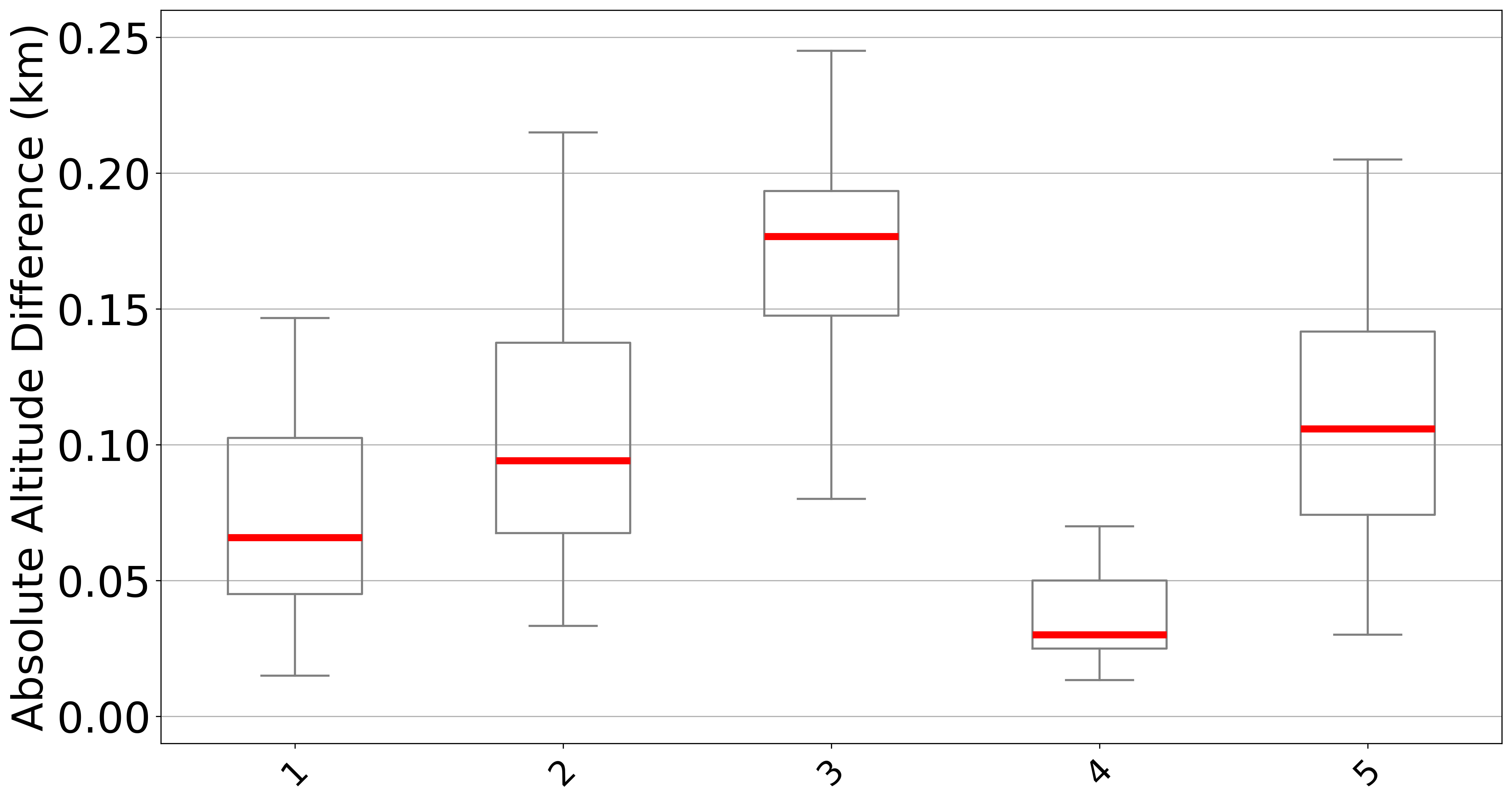}
        \caption[]%
        {{\small 97.6$\degree$ Inclination Groups}%
        }
        \label{fig:97_inclination_appendix}
    \end{subfigure}
    \caption{\small Altitude changes across orbital groups at four inclinations (43°, 53°, 70°, and 97.6°) during the May 2024 superstorm peak. Box plots show significant variation within inclinations and distinctive "W" patterns indicating systematic impact based on satellite orientation relative to solar storm direction.}
    \label{fig:inclinations_plot_appendix}
\end{figure*}

\begin{figure*}
    \centering
    \begin{subfigure}{0.48\textwidth}
        \includegraphics[width=0.9\textwidth]{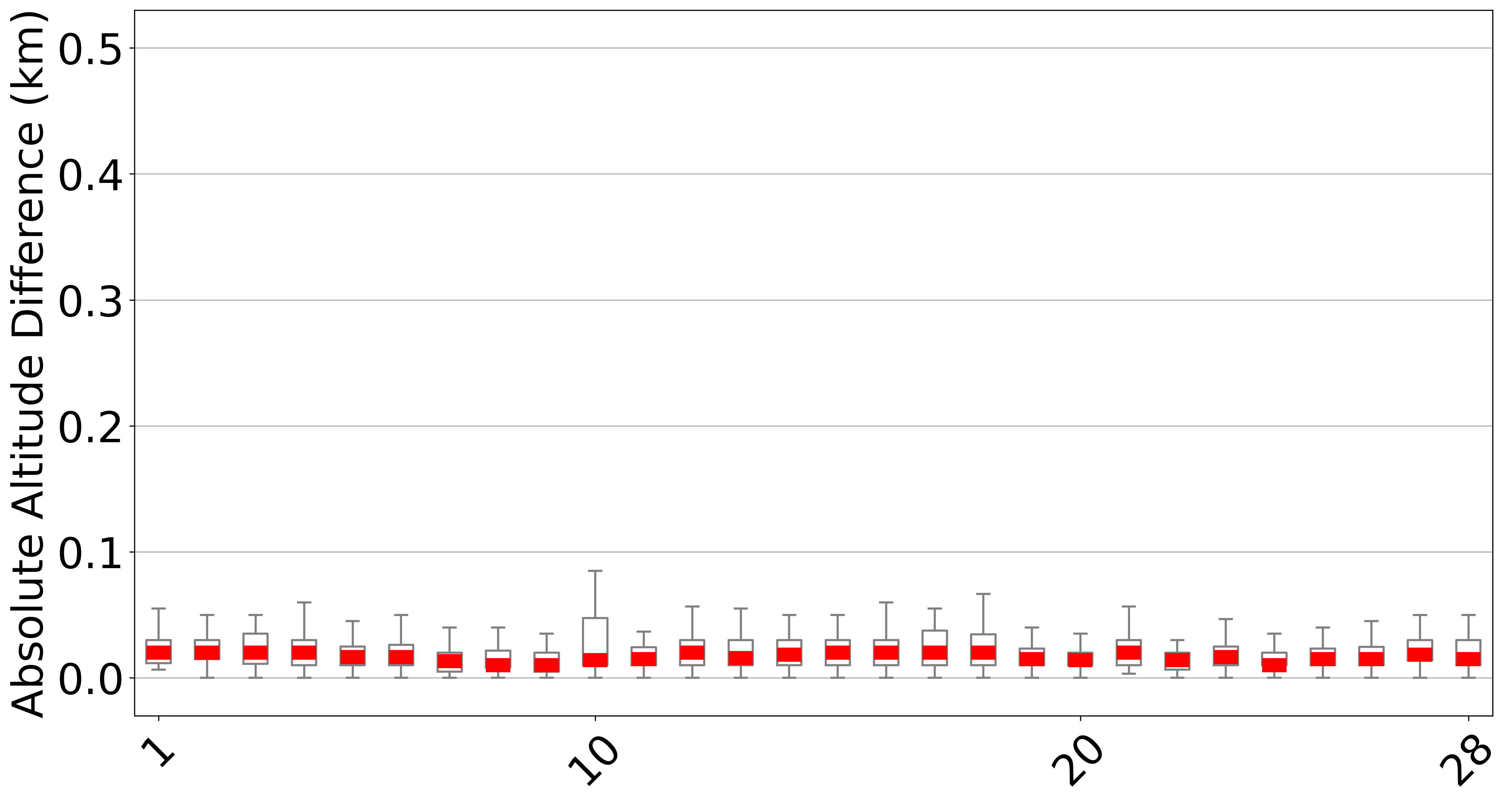}
        \caption[]%
        {{\small 43$\degree$ Inclination Groups}%
        }
        \label{fig:43_inclination_appendix}
    \end{subfigure}
    \hfill
    \begin{subfigure}{0.48\textwidth}
        \includegraphics[width=0.9\textwidth]{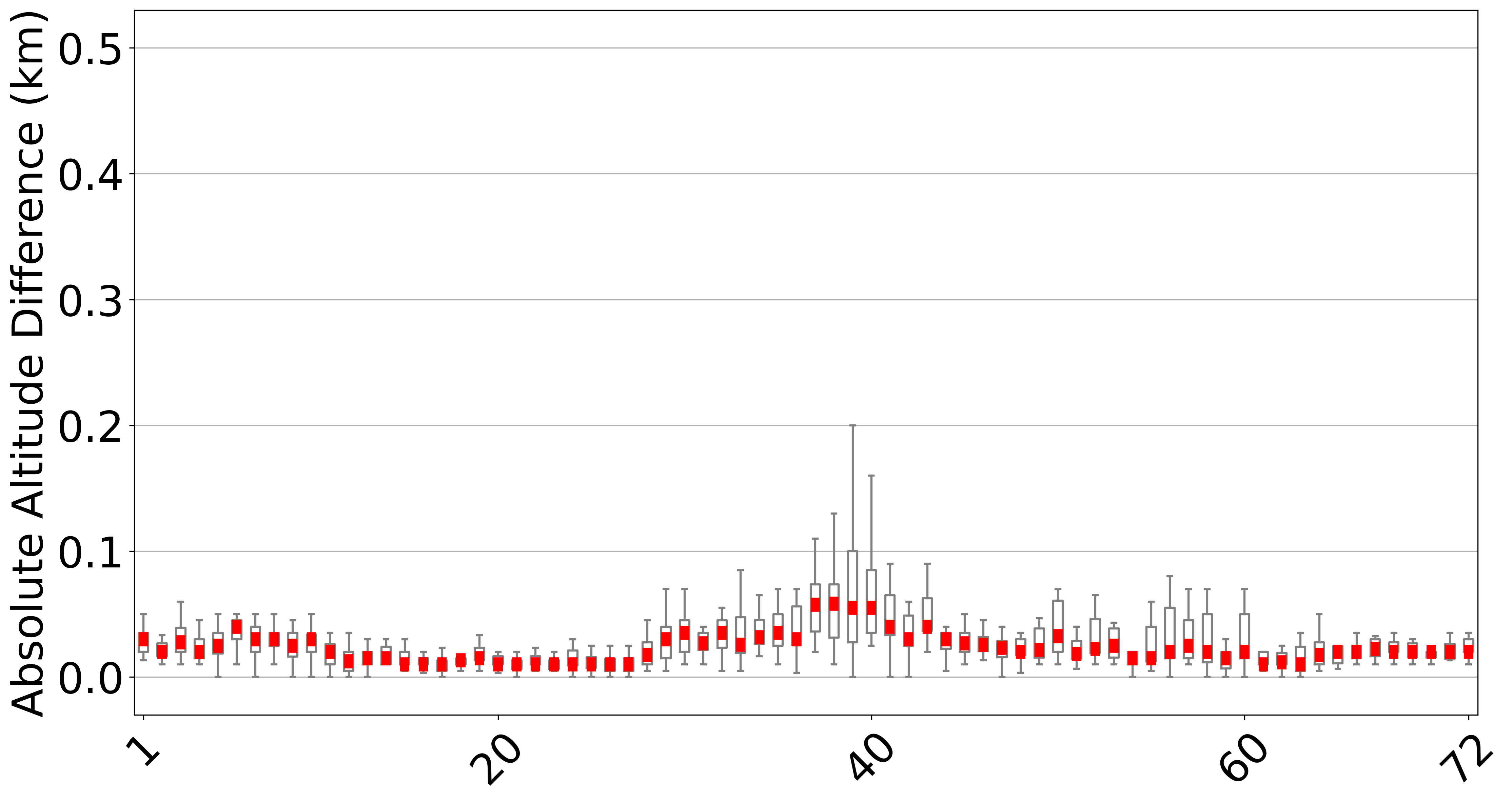}
        \caption[]%
        {{\small 53$\degree$ Inclination Groups}%
        }
        \label{fig:53_inclination_appendix}
    \end{subfigure}
    \medskip
    \begin{subfigure}{0.48\textwidth}
        \includegraphics[width=0.9\textwidth]{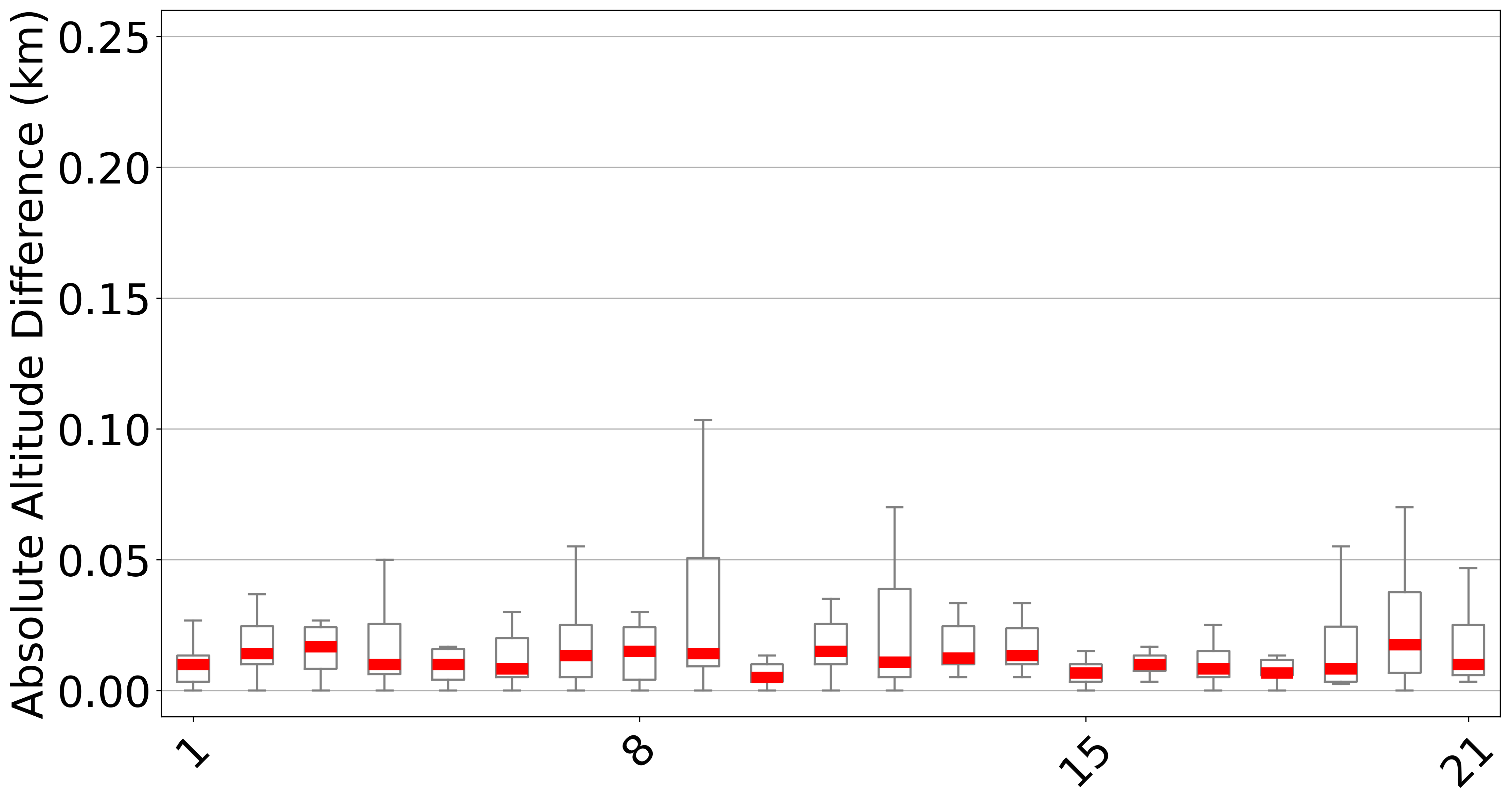}
        \caption[]%
        {{\small 70$\degree$ Inclination Groups}%
        }
        \label{fig:70_inclination_appendix}
    \end{subfigure}
    \hfill
    \begin{subfigure}{0.48\textwidth}
        \includegraphics[width=0.9\textwidth]{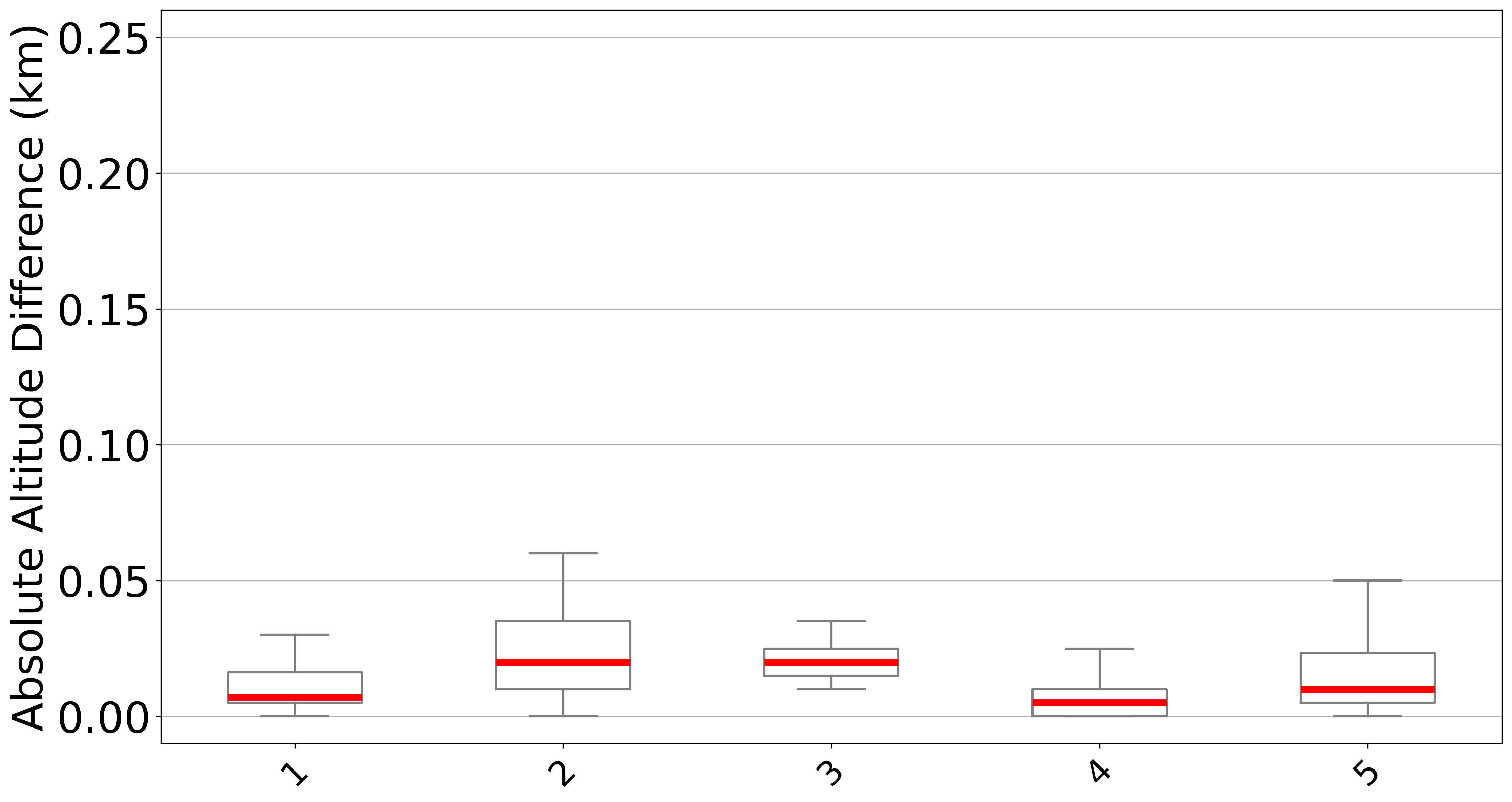}
        \caption[]%
        {{\small 97.6$\degree$ Inclination Groups}%
        }
        \label{fig:97_inclination_appendix}
    \end{subfigure}
    \caption{\small Altitude changes across orbital groups at four inclinations (43°, 53°, 70°, and 97.6°) during a quiet period (April 11\textsuperscript{th}). The altitude changes are minimal and nearly uniform across all orbits. The y-axis of the plot is set to the same range as Figure~\ref{fig:inclinations_plot_appendix} to enable visual comparison.}
    \label{fig:inclinations_plot_baseline_ylim_appendix}
\end{figure*}

\section{Appendix}
Figure~\ref{fig:inclinations_plot_appendix} presents box plots of altitude changes of the satellite across orbits at three inclinations (43°, 53°, 70°, and 97.6°) during the May 2024 superstorm peak (May 11\textsuperscript{th}). We observe the distinctive ``W'' pattern across all inclinations. Figure~\ref{fig:inclinations_plot_baseline_ylim_appendix} shows the altitude changes during a quiet period on April 11\textsuperscript{th} and serves as the baseline for comparing storm and non-storm periods. The baseline period in Figure~\ref{fig:inclinations_plot_baseline_ylim_appendix} has nearly uniform changes across orbit groups, while the characteristic ``W'' pattern in Figure~\ref{fig:inclinations_plot_appendix} reflects the impact of the solar storm during the storm period.

\end{document}